\documentclass[a4paper,11pt]{article}

\usepackage{jinstpub} 

\usepackage{lineno}
\usepackage{hyperref}
\usepackage[figure]{hypcap}

\usepackage{upgreek}

\usepackage{placeins}

\title{\boldmath Development of a Bi-solvent Liquid Scintillator with Slow Light Emission}

\author[a,b,c,1]{Hans Th. J. Steiger,\note{Corresponding author.}}
\author[a]{Matthias Raphael Stock,}
\author[b,c]{Manuel Böhles,}
\author[a]{Sarah Braun,}
\author[d,e]{Edward J. Callaghan,}
\author[a]{David Dörflinger,}
\author[a]{Ulrike Fahrendholz,}
\author[a]{Jonas Firsching,}
\author[a]{Elias Fischer,}
\author[d,e]{Tanner Kaptanoglu,}
\author[a]{Lennard Kayser,}
\author[a]{Meishu Lu,}
\author[a]{Lothar Oberauer,}
\author[d,e]{Gabriel D. Orebi Gann,}
\author[a]{Korbinian Stangler,}
\author[b,c]{Michael Wurm,}
\author[b,c]{Dorina Zundel}

\affiliation[a]{Technical University of Munich, TUM School of Natural Sciences, Physics Department, \\ James-Franck-Str. 1, 85748 Garching, Germany}

\affiliation[b]{Cluster of Excellence PRISMA$^+$ \\Staudingerweg 9, 55128 Mainz, Germany}
\affiliation[c]{Institute for Physics, Johannes Gutenberg University Mainz \\ Staudingerweg 7, 55128 Mainz, Germany}

\affiliation[d]{Lawrence Berkeley National Laboratory, \\ 1 Cyclotron Road, Berkeley, CA 94720-8153, USA}
\affiliation[e]{Physics Department, University of California at Berkeley, \\ 366 Physics North MC 7300, Berkeley, CA 94720-7300}
\emailAdd{hans.steiger@tum.de}

\abstract{One of the most promising approaches for the next generation of neutrino experiments is the realization of large hybrid Cherenkov/scintillation detectors made possible by recent innovations in photodetection technology and liquid scintillator chemistry. The development of a potentially suitable future detector liquid with particularly slow light emission is discussed in the present publication. This cocktail is compared with respect to its fundamental characteristics (scintillation efficiency, transparency, and time profile of light emission) with liquid scintillators currently used in large-scale neutrino detectors. In addition, the optimization of the admixture of wavelength shifters for a scintillator with particularly high light emission is presented. Furthermore, the pulse-shape discrimination capabilities of the novel medium was studied using a pulsed particle accelerator driven neutron source. Beyond that, purification methods based on column chromatography and fractional vacuum distillation for the co-solvent DIN (Diisopropylnaphthalene) are discussed.}


\arxivnumber{1234.56789} 

\begin{document}
\maketitle
\flushbottom

\section{Introduction}

One of the most promising approaches for the next generation of neutrino experiments is the realization of large hybrid Cherenkov/scintillation~(C/S) detectors. The potential arising from time separation of the components of the light signal is foreseen to be realized with innovations in photon detection technology and liquid scintillator chemistry. This publication is focused on slow scintillators~(see~\cite{Biller1, Aberle1, Li1, Wei1, Guo1}). In such a hybrid detector it would be possible to exploit the Cherenkov signal for the reconstruction of directional and topological information while the high light yield of an organic scintillator would ensure the good energy resolution and low thresholds necessary for several applications~\cite{Theia:2019non}. From the ratio of Cherenkov to scintillation signal strength particle identification and background suppression potential arises in addition to the well known pulse-shape discrimination capabilities of most liquid scintillators used in neutrino physics~\cite{Biller1, Biller2}.

There are two alternative approaches followed up for the hybrid technique: in the context of long baseline oscillation experiments, Water-based Liquid Scintillator~(WbLS) has been developed as a weakly scintillating and highly transparent target material~\cite{Yeh1, Tanner1}. While offering attractive properties and a cost-efficient way to realize very large detectors aiming for signals in the hundreds of~MeV to~GeV range,~WbLS in general offers relatively low scintillation light yield and fast scintillation times. This limits the achievable lower energy threshold and resolution and requires either very fast light sensors~(e.g.~LAPPDs), wavelength-sorting~(e.g. dichroicons \cite{dichroicon}) or a high granularity of photo sensors to distinguish Cherenkov and scintillation signals \cite{Tanner2}. The applicability of~WbLS in the~MeV to~GeV range is currently investigated in a number of ton-scale demonstrator \mbox{detectors \cite{ANNIE:2023yny,Anderson:2022lbb,Zhao:2023ydx}}.

However, in the detection of low-energy astrophysical neutrinos or the search for neutrino-less double beta decay, the typical high light yields of conventional organic scintillators are absolutely necessary to achieve sub-MeV detection thresholds and per cent-level energy resolutions \cite{Dunger:2022gif}. In this context, so-called \emph{slow scintillators} are a very attractive alternative for hybrid detection. \emph{Slow scintillators} are mostly organic and not as transparent as~WbLS. Instead, they permit for a separate detection of Cherenkov and scintillation signals by slowing down the scintillation light emission and effectively delaying the signal by 10 nanoseconds or more. In this way, C/S separation becomes possible despite of the disproportionately large scintillation signal and using conventional PMTs only. In principle, the required delay in scintillation times is straight-forward to achieve by reducing the fluor concentration in the cocktail and thus the efficiency of the excitation transfer from solvent to fluor~\cite{Guo1, Yeh2}). However, this loss in transfer efficiency is accompanied with a strong reduction in scintillation light yield. In Ref.~\cite{Biller2}, S.~Biller {\it et al.}~were able to show that organic scintillators can be slowed down by use of fluors selected for long emission times, suffering only moderate loss of total light yield. 
While a promising alternative, these fluors~(e.g.~9.10-Diphenylanthracene,~CAS 1499-10-1) are not yet mass-produced in the required purity and therefore, still expensive. Moreover, the radiochemical purification of these agents is still the subject of current research. Likewise, it has not yet been shown that such a scintillator mixture allows~PID through the pulse-shape of the scintillation.

Instead, here we present a wholy novel approach to create a slow organic scintillator by blending two solvents, linear alkylbenzene~(LAB, CAS~67774-74-7) and Diisopropylnaphthalene~(DIN, CAS~38640-62-9). This basic mixture is then doped with regular fluors,~e.g.~the commonly used PPO~(2,5-Diphenyloxazole,~CAS~92-71-7) or slightly rarer~BPO~(2-(4-Biphenyl)-5-phenyloxazole,~CAS~852-37-9). The resulting cocktails show the desired slow light emission but at the same time are able to maintain light yields and pulse-shape capabilities fully comparable to a regular organic scintillator. This makes these bi-solvent scintillators especially attractive for applications where high energy resolution and efficient background suppression are of paramount importance, especially for the next generation of neutrino-less double-beta decay experiments with loaded scintillators \cite{Dunger:2022gif,Theia:2019non}. Note that these applications also require high optical coverage, so that a detector equipped with conventional PMTs or a scintillator upgrade of an existing detector such as~KamLAND,~SNO+ or~JUNO would be very cost-effective. It should be mentioned here, that by slowing down the scintillation light emission by this method as well as following the approach of S.~Biller {\it et al.} using slow fluors ~\cite{Biller2}, directional information can be gained effectively from Cherenkov photons also for low energy depositions in the scintillation medium. Nonetheless, the vertex resolution might be degraded to some extent by the slow light emission, making a careful tuning with respect to the~C/S-separation quality necessary.  

In the present paper, we report both the recipe and the characteristics of the new bi-solvent \emph{slow scintillator}. We investigate its light yield, transparency, spectral light emission and characteristic pulse-shape for e$^-$/$\gamma$-interactions and neutron induced proton recoils. Beyond that, purification techniques based on column chromatography and fractional vacuum distillation for the co-solvent~DIN are presented.  


\section{Composition and production}
\label{sec:production}

As previously mentioned, the investigated slow scintillation mixtures contain~LAB as main solvent. The solvent used here was produced by~SINOPEC Jinling Petrochemical Company and purified during the~JUNO liquid scintillator pilot plant commissioning phase at the Daya Bay Neutrino Laboratory. Besides~Al$_2$O$_3$ column chromatography, also fractional vacuum distillation was applied. The~LAB is a mixture of compounds that can be expressed in terms of~n in the form of ${(\text{C}_6\text{H}_5)-\text{C}_\text{n}\text{H}_{2\text{n}+1}}$ with $\text{n}=10-13$ (for details on molecular composition and applied purification techniques see \cite{Lombardi1}). An attenuation length of~$\Lambda_{\text{LAB}}(430\,\text{nm})$~=~(28.07~$\pm$~2.94)\,m for light with a wavelength of~430\,nm was achieved after the vacuum distillation step~\cite{Lombardi1, Franke}.\\
The second solvent blended~(in concentrations of~10\,vol\% and~20\,vol\%) with the~LAB is a mixture of isomers of Diisopropylnaphtalin, which was purchased from Rütgers~(Rain Carbon Inc.). The used product is specified as "scintillation grade" and sold under the brand name RUETASOLV~DI-S. While the flashpoints of~DIN and~LAB are similar~$\sim$140$^{\circ}$C the density of the DI-S 0.9450\,$-$\,0.9700\,\text{g/cm}$^3$~\cite{RAIN} is slightly higher than for LAB ($\sim$0.87\,\text{g/cm}$^3$). Also the refractive index of~$\sim$1.57~\cite{RAIN} exceeds the one of LAB~($\sim$1.49). In general, the transparency of DIN has so far proved to be critical. Even the DIN used here, as well as commercial DIN based scintillators~(e.g.~EJ-309~\cite{EJ-309}) do not always achieve attenuation lengths above~1\,m. As a consequence, the RUETASOLV~DI-S had to be purified by~{Al$_2$O$_3$} column chromatography and fractional vacuum distillation to enhance its optical properties for further sample production.

\subsection{Purification of DIN}

\begin{figure}[h]
    \centering
    \includegraphics[width=1.00\textwidth]{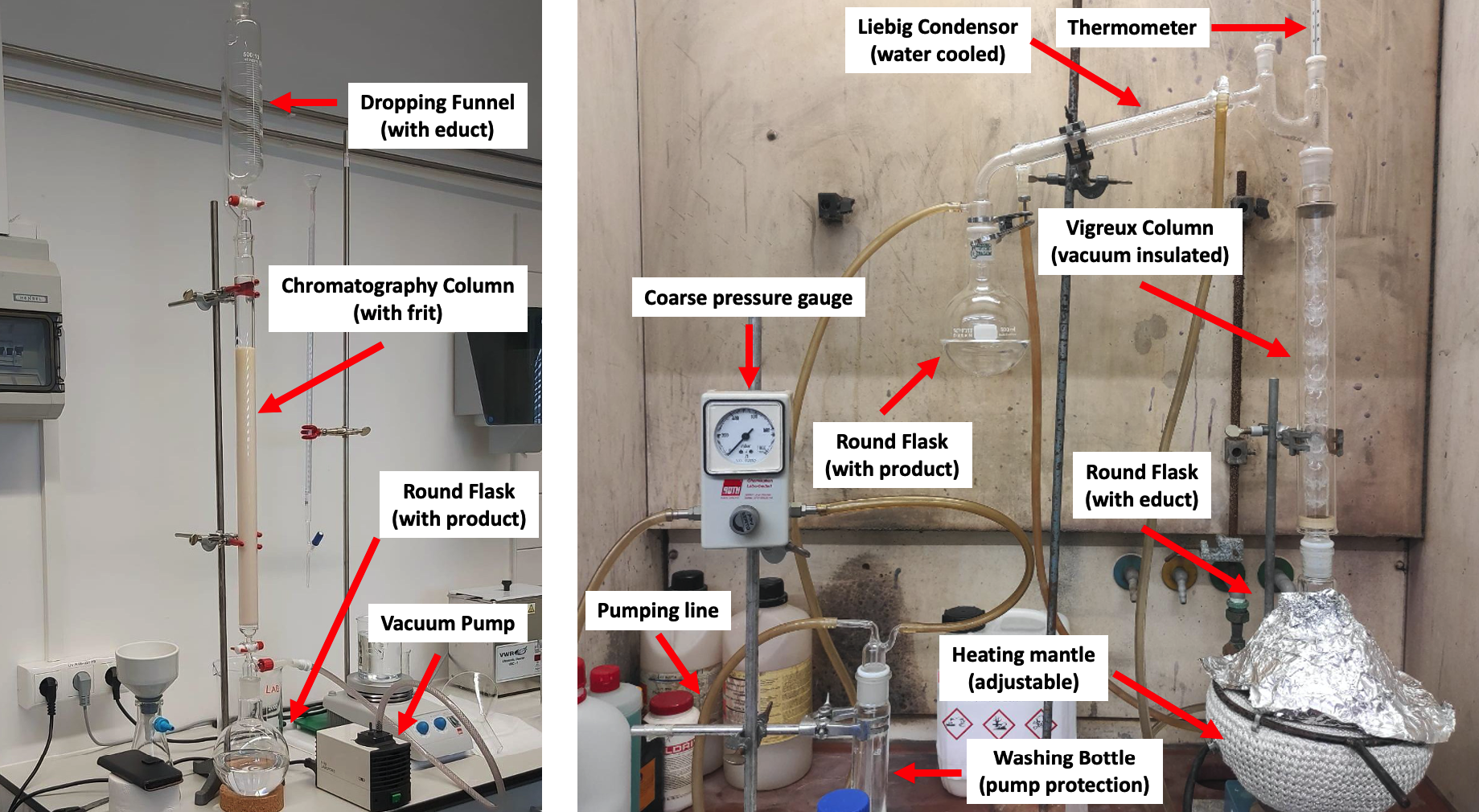}
    \caption{\textbf{Left:} Chromatography setup as used for the purification of~RUETASOLV~DI-S. By means of a vacuum pump the filtration process was accelerated to a flow rate of~$\sim$350\,ml/h while carefully maintaining the temperature of the column close to the ambient one. \textbf{Right:} Distillation apparatus for the fractional vacuum distillation of DIN. The Vigreux column with insulating vacuum jacket helps to improve the thermal separation of products and educts.}
    \label{fig:Purification}
\end{figure}

In a first purification step~(shown in~\autoref{fig:Purification} left) the~DIN was purified using a~600\,mm long chromatography column~(inner diameter:~30\,mm) filled with basic~{Al$_2$O$_3$} powder~(CAS~1344-28-1) in the activity stage~I~(0\%~water) and~pH~$= 8.5$. The column contained a frit to retain the alumina powder. Adding the DIN very slowly into the column by careful operation of the dropping funnel prevented excessive heating and thus degradation of the purification performance of the alumina substrate. After completion of the chromatography process, clear brownish and yellowish stripes were visible in the stationary phase of the column. Subsequent filtration of the~DIN was performed, using a Büchner funnel equipped with ashless filter paper~(retention:~2\,\text{$\upmu$}m) to remove any un-retained alumina crystals that might have passed through the frit. 

In a second step purification by vacuum distillation was applied to the DIN. Therefore, the apparatus shown in~\autoref{fig:Purification}~(right) was constructed. To enhance the separation of the raw DIN and the distillation products a Vigreux column with an effective length of~40\,cm and an vacuum insulation jacket was mounted. For condensing the purified DIN a conventional water cooled Liebig Condensor is used. To protect the DIN vapor from oxygen the apparatus can be flushed with dried nitrogen while constantly being pumped down to a pressure below~$\sim$10\,mbar. The temperature~(typically~$120^{\circ}$C\,$-$\,140$^{\circ}$C) is adjusted such, that the (very similar) boiling points of all DIN-isomers are well exceeded. After distillation is complete, the temperature in the apparatus is slowly reduced to ambient conditions before the vacuum is broken. The purified DIN is then purged extensively with dried nitrogen gas to saturate it with an inert gas. The residue remaining in the educt flask has a viscous consistency and an intensive yellowish to brownish color similar to the one observed in the chromatography column.

\subsection{Transparency of DIN and LAB/DIN-Mixtures}

\begin{figure}[h]
    \centering
    \includegraphics[width=0.75\textwidth]{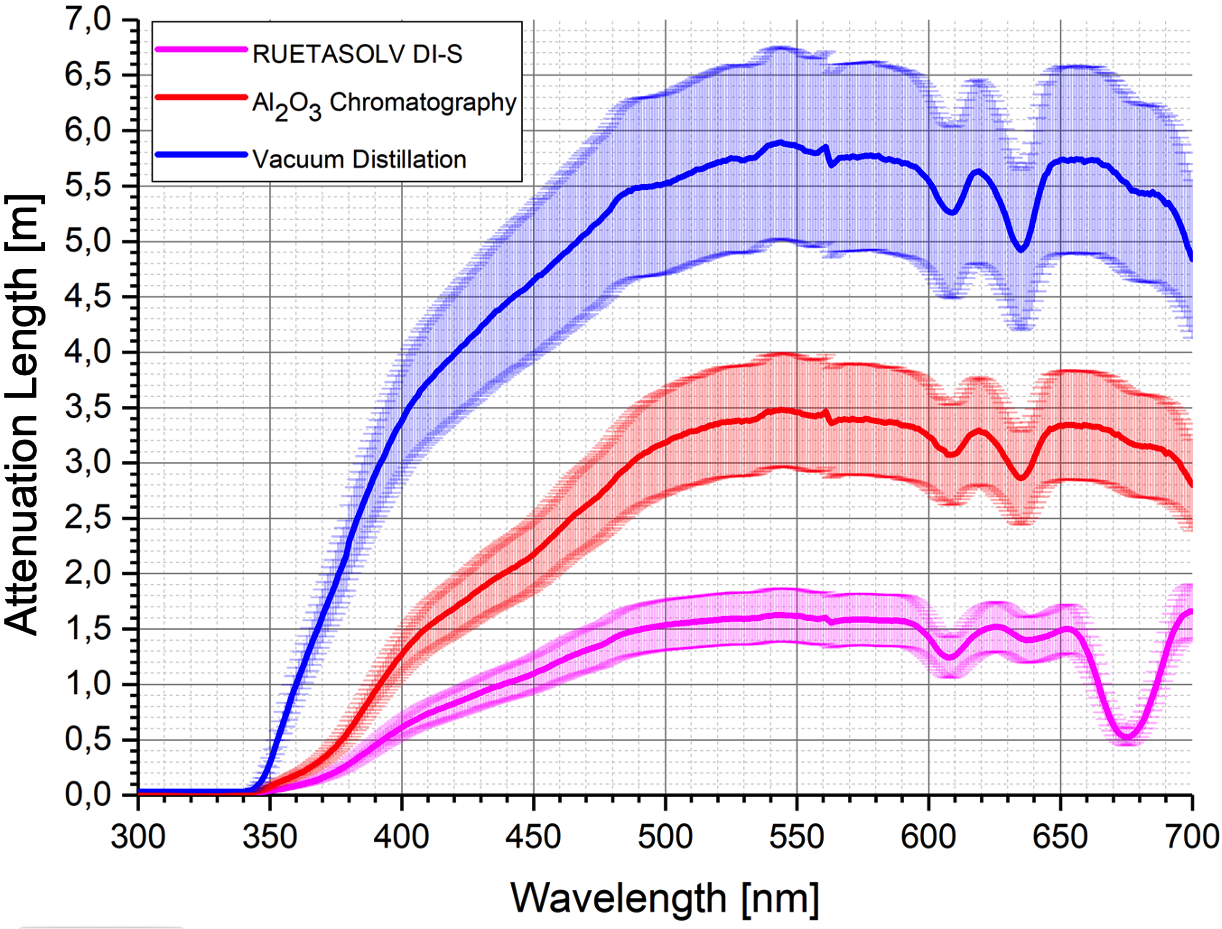}
    \caption{\textbf{Left:} Attenuation length measurement of the co-solvent RUETASOLV DI-S unpurified (purple), after chromatography (red) and vacuum distilled (blue) from~300\,nm to~700\,nm of wavelength.}
    \label{fig:UVVis}
\end{figure}

In order to evaluate the transparency and thus the purification achieved during the previously shown methods, the attenuation length of the raw DIN, as well as after column chromatography and distillation was measured. Therefore, a Perkin Elmer Lambda~850+~UV/Vis-spectrometer with a long rectangular fused-silica cuvette~($L~=~10$\,cm optical path length) was used to gain the results shown in \autoref{fig:UVVis}.        

Here, the improvement of the transmission in the range of~350\,nm\,$-$\,450\,nm is of particular interest, since this is where most fluors emit most of their light and the spectral quantum efficiency of the photosensors commonly used in neutrino physics reaches its maximum. The extinction coefficient of most fluors and with that their absorbance become negligible above~$\sim$400\,nm~\cite{AberleJINST}. In this wavelength region the impurity levels of the used chemicals and scattering processes dominate the light attenuation. Therefore, a comparison of the attenuation length at~430\,nm has become common in literature. For the raw RUETASOLV~DI-S an attenuation length of~$\Lambda_{\text{raw}}(430\,\text{nm})=(0.92~\pm~0.14)$\,m was determined. As already observed by Song et al.~\cite{Song} the column chromatography with~Al$_2$O$_3$ improves the transparency only slightly. A value of~$\Lambda_{\text{Al}_\text{2}\text{O}_\text{3}}(430\,\text{nm})=(1.87~\pm~0.28)$\,m is still problematic for a usage in multi-ton scale detectors. However, fractional vacuum distillation with the Vigreux column of the previously chromatographed DIN provided a greater improvement in attenuation length to~$\Lambda_{\text{dist.}}(430\,\text{nm})=(4.23\pm0.62)$\,m. The attenuation length of the mixtures of the highly transparent LAB and the purified DIN can be calculated via

\begin{equation}
\frac{1}{\Lambda}=\sum_i\frac{1}{\Lambda_i}
\end{equation}

as the total absorbance~$\Lambda$ of a mixture of liquids can be calculated by adding up the individual contributions~$\Lambda_i$. As mentioned above the extinction at the wavelength region of interest is dominated by impurities in the liquids, which can be assumed to be very small compared to the solvent's concentration~\cite{BuckYeh}. Therefore, and given the similar molar masses of LAB and DIN, linearity between the mass concentration of the individual solvents and their absorbance can be assumed. For a mixture of 90\,vol\% LAB and~10\,vol\% of the fully purified~DIN the attenuation length 

\begin{equation}
\Lambda_{90/10}(430\,\text{nm})=~(17.5~\pm~2.3)\,\text{m}
\end{equation}
at~430\,nm can be estimated. Assuming that the previous assumptions still hold for a~20\,vol\% admixture of~DIN, the attenuation length is 

\begin{equation}
\Lambda_{80/20}(430\,\text{nm})=~(13.9~\pm~1.6)\,\text{m}\,.
\end{equation}

\subsection{Produced Scintillator Samples}
\enlargethispage{1cm}
Liquid scintillators based on the so called classical or low flashpoint solvents~(e.g. Xylene or Toluene) typically show a very short pulse length. In commercial cocktails optimized for~PSD like~{EJ-301}~(Eljen
Technology),~{BC-501A}~(Bicron, Saint Gobain) or~{NE-213}~(Nuclear Enterprises) the admixture of Naphtalene in concentrations of~$\approx$8\%\,$-$\,10\% stretches the light emission considerably. Furthermore, Naphtalene enhances the~PSD capabilities~\cite{Perkin, Bonhomme}. In addition to limited transparency of such high-percentage naphthalene solutions, the durability of these scintillators is known to be limited. Among other things, the naphthalene (which is a crystaline powder at room temperature) begins to crystallize on the detector vessel walls over time, slowly changing basic properties of the mixture. On the other hand significant evaporation of the volatile xylene can also change the cocktails composition~\cite{Bonhomme}. Furthermore, their low flashpoint makes these scintillators not suitable for applications with strict safety requirements as large scale neutrino detectors.

As~DIN itself is a derivative of naphthalene a similar effect on the scintillation time constants was expected while mixtures of liquid~DIN and other solvents would not show the mentioned instability issue due to crystallization or volatility. Since the aim of this work is to produce transparent scintillators with a high flashpoint and low environmental hazard, whose light emission is particularly slow, the choice of the primary solvent LAB was inevitable. In order to study the effects of DIN admixtures, the samples as listed in~\autoref{Chems} were synthesized.

\begin{table}[h]
\centering
\begin{tabular}{|c|c|c|c|c|}
\hline 
Sample Name & Solvent & Co-Solvent & Fluor & Fluor Concentration \\ 
\hline 
PPO-05 & 90\% LAB & 10\% DIN & PPO & 0.5\,g/l \\
\hline 
PPO-10 & 90\% LAB & 10\% DIN & PPO & 1.0\,g/l \\ 
\hline 
PPO-20 & 90\% LAB & 10\% DIN & PPO & 2.0\,g/l \\ 
\hline 
BPO-05 & 90\% LAB & 10\% DIN & BPO & 0.5\,g/l \\ 
\hline 
BPO-10 & 90\% LAB & 10\% DIN & BPO & 1.0\,g/l \\ 
\hline 
BPO-20 & 90\% LAB & 10\% DIN & BPO & 2.0\,g/l \\ 
\hline 
PPO-05-20 & 80\% LAB & 20\% DIN & PPO & 0.5\,g/l \\ 
\hline 
PPO-10-20 & 80\% LAB & 20\% DIN & PPO & 1.0\,g/l \\ 
\hline 
PPO-20-20 & 80\% LAB & 20\% DIN & PPO & 2.0\,g/l \\ 
\hline 
\end{tabular} 
\caption{The composition of the produced slow LS samples with blended solvents. The main component LAB is used in concentrations of~90\% and~80\% by volume respectively. DIN was added as co-solvent. Six samples use the most commonly used fluor~PPO in neutrino physics, while the samples containing~BPO where synthesized to optimize the energy transfer from DIN to the fluor. The samples~{PPO-05},~{PPO-10},~{PPO-20} and~{BPO-10} where completely characterized in the scope of this work, while on the samples only light yield measurements (for comparison) where performed.}
\label{Chems}
\end{table}

PPO was used as a fluor because it is widely used in large scale monolithic neutrino detectors with excellent purity and price. For some samples containing a~90/10 mixture of the solvents the energy transfer between DIN and the fluor was optimized by the usage of~BPO, which is known to provide high light yields in DIN-based cocktails~\cite{BuckYeh}. Both fluors were provided by Sigma Aldrich in the quality "scintillation grade" and used without further purification. All samples were prepared under exclusion of air and permanent bubbling of dried nitrogen through the liquid in the round mixing flask used. After the fluors have dissolved completely, the scintillators were purged with nitrogen for more than~30\,min and stored in commercial Duran bottles under protective atmosphere. Before filling test detectors with the samples, they were again purged with nitrogen and sealed in the respective vessel under moderate overpressure of the inert~N$_2$ gas.

\section{Scintillation properties}
\label{sec:scintillation}

\subsection{Spectral Scintillation Light Emission}

\begin{figure}[h!]
    \centering
    \includegraphics[width=0.495\textwidth]{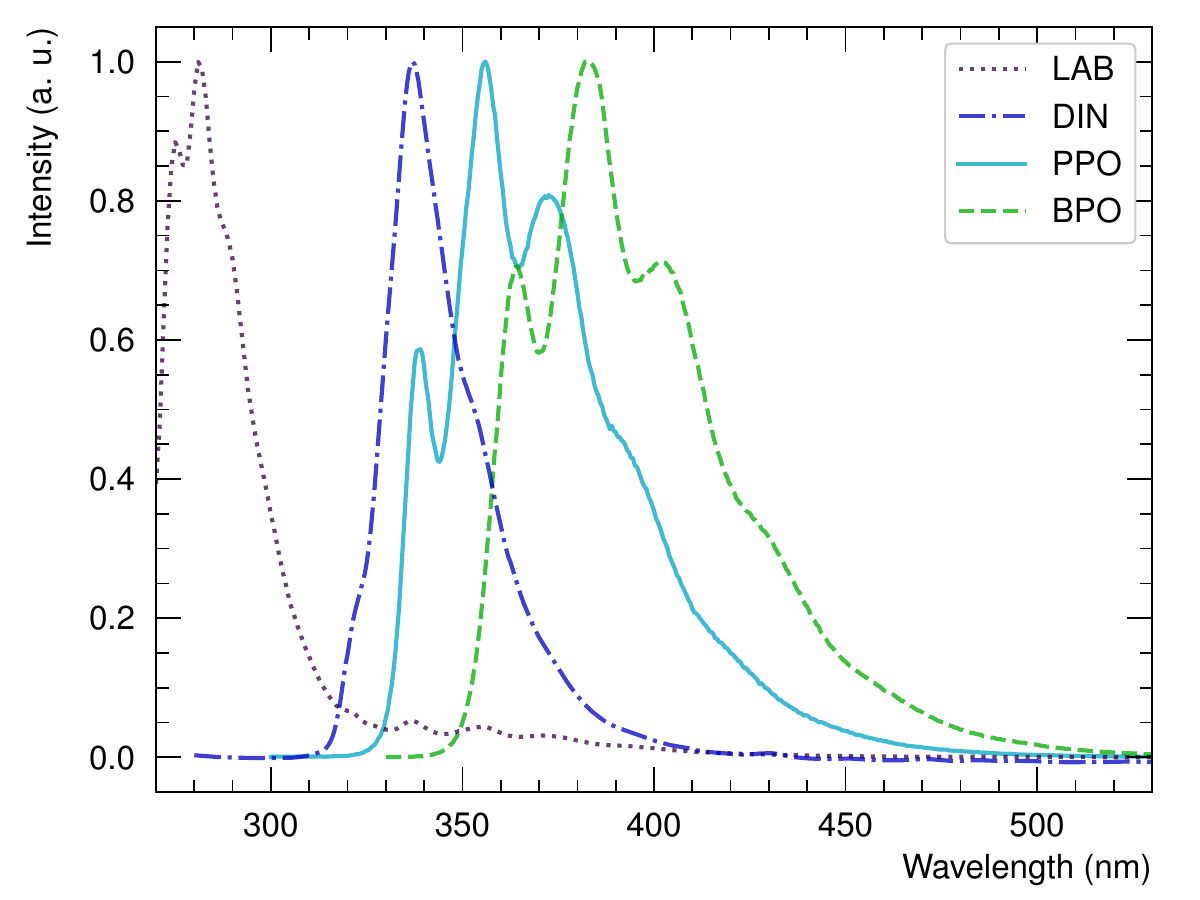}
    \includegraphics[width=0.495\textwidth]{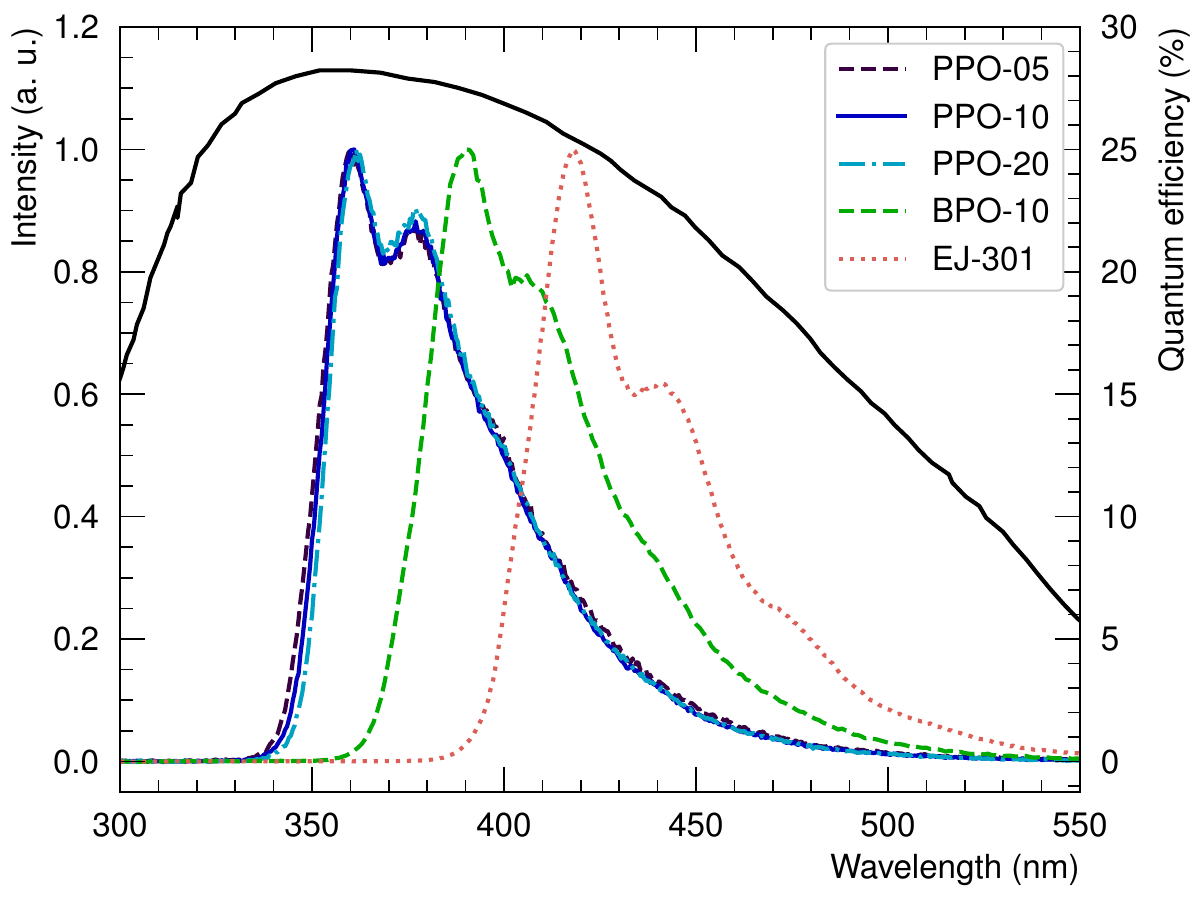}
    \caption{\textbf{Left:} Emission spectra of the substances used in the sample cocktails normalized to their global maxima. To minimize the self-absorption and re-emission of the substances, they were strongly diluted in cyclohexane~(CHX). The spectra were recorded in front-face geometry. \textbf{Right:} Emission spectra for three samples containing~PPO and one~BPO. As a reference the emission curve of~{EJ-301} which served as a standard in later light yield measurements is depicted~(red). To evaluate the light emission as seen by~PMTs in detector application all spectra were recorded in conventional geometry resulting in long light paths and self-absorption in the liquids. All samples were undiluted. The black line represents the spectral quantum efficiency of the conventional~PMTs~(ETEL~9128B) used in the setup to determine the light yield of the liquid scintillator candidates~(see also~\autoref{fig:LySetupMZ}) as given in the datasheet~\cite{ETEL9128B}.}
    \label{fig:SpecFluo}
\end{figure}

The wavelength-dependent emission of the individual sample components (solvents and fluors) as well as the effective emission spectrum of the scintillator mixtures were evaluated using an Edinburgh~FS5 spectro-fluorometer. To minimize the effect of absorption and re-emission of the pure scintillator components they were diluted in cyclohexane~(CAS 110-82-7,~HPLC grade) and measured using a front-face geometry holder (Edinburgh SC-10) to limit the optical path length through the medium itself (see~\autoref{fig:SpecFluo} left and \autoref{FluoTab}).

\begin{table}[h!]
\centering
\begin{small}
\begin{tabular}{|c|c|c|c|c|c|}
\hline 
Substance & Chemical Formula & Conc. in CHX & Exc.~[nm] & Abs. Max.~[nm] & Em. Max.~[nm] \\ 
\hline 
LAB & ${(\text{C}_6\text{H}_5)-\text{C}_\text{n}\text{H}_{2\text{n}+1}}$ & 10~ml/l & 255 $\pm$ 1 & 260~\cite{BuckYeh} & 281 $\pm$ 1 \\
 & with $\text{n}=10-13$ & & & & \\ 
\hline 
DIN & C$_{16}$H$_{20}$ & 10\,ml/l & 255 $\pm$ 1 & 279~\cite{BuckYeh} & 337 $\pm$ 1\\ 
\hline 
PPO & C$_{15}$H$_{11}$NO & 40\,mg/l & 290 $\pm$ 1 & 303~\cite{BuckYeh} & 356 $\pm$ 1 \\ 
\hline 
BPO & C$_{21}$H$_{15}$NO & 100\,mg/l & 320 $\pm$ 1 & 320~\cite{BuckYeh} & 383 $\pm$ 1\\ 
\hline 
\end{tabular} 
\end{small}
\caption{Listed are the absorption~(Abs. Max.) and emission~(Em. Max.) maxima for the chemical compounds used in the scintillator samples. All chemicals were diluted in cyclohexane and studied in a front face geometry arrangement to limit self-absorption and re-emission in the cuvette.} 
\label{FluoTab}
\end{table}

The emission of our~LAB peaking at~(281~$\pm$~1)\,nm matches well the absorption spectrum of the second solvent~DIN peaking at~279\,nm~\cite{BuckYeh}. The emission spectrum of~DIN is shifted towards higher wavelengths (maximum at~(337~$\pm$~1)\,nm). Therefore, the overlap with the~PPO absorption spectrum is not ideal. In contrast,~BPO with its absorption maximum at~320\,nm~\cite{BuckYeh} matches better the~DIN's emission and thus leads to an enhanced energy transfer.   

To evaluate the emission spectrum of the scintillator mixtures (as seen by photosensors in a detector), the samples~{PPO-05},~{PPO-10},~{PPO-20} and~{BPO-10} were examined and compared with that of the commercial cocktail~{EJ-301}. To take into account the self-absorption of the samples, they were measured pure and undiluted in a $10\times10\times40$\,mm$^3$ fused silica cuvette. The conventional geometry sample holder~(Edinburgh SC-05) ensures a long light path through the sample. All scintillators were excited close to the~LAB's maximum absorption with light of~(255~$\pm$~1)\,nm wavelength, where~0.5\,nm was selected as emission and~1\,nm excitation slit-widths. The resulting spectra are depicted in the right plot of~\autoref{fig:SpecFluo}. All samples containing~PPO show, as expected from the similar fluor concentrations, almost the same spectrum, which reaches its maximum at~$\sim$361\,nm~({PPO-05}:~(360~$\pm$~1)\,nm,~{PPO-10}:~(361~$\pm$~1)\,nm,~PPO-20:~(362~$\pm$~1)\,nm). Only a slight shift in the spectrum to higher wavelengths for larger amounts of fluor can be seen as expected. This effect is fully explained by the increased self-absorption. Even though the spectrum of these scintillators is still well within the ultraviolet range, where typical detector materials such as acrylics do not yet reach their full transparency, the~PPO preserves completely the visible part of the Cherenkov spectrum as it lies above the fluor's absorption range. Furthermore, the PPO emission spectrum matches already well the spectral quantum efficiency of conventional bi-alkali~PMTs. The emission spectrum of~{BPO-10} is peaking at~(390~$\pm$~1)\,nm with a tail into the visible regime up to~$\sim$500\,nm. Therefore, an even better match of~PMT's~QE spectrum and transmission of detector vessels can be expected, while the transparency of the scintillator for the Cherenkov component around $\sim$400\,$-$\,420\,nm is reduced compared to~PPO. As a reference and because it was used as a standard in later light yield measurements, the known spectrum of~{EJ-301} was also measured (see red curve in~\autoref{fig:SpecFluo}). The scintillator, which already appears to the naked eye as strongly blue shining, reaches its emission maximum at~(418~$\pm$~1)\,nm which is caused by the addition of secondary wavelength shifters with corresponding emission spectra in such commercial cocktails. It should be mentioned here, that the use of secondary wavelength shifters~(WLS) should be avoided as much as possible in scintillators optimized with regard to high transparency for their own Cherenkov light. The additional absorption band introduced by the secondary~WLS cuts away additional parts of the spectrum. 


\FloatBarrier

\subsection{Light yield}
\label{subsec:lightyield}

\begin{figure}[h!]
    \centering
    \includegraphics[width=0.50\textwidth]{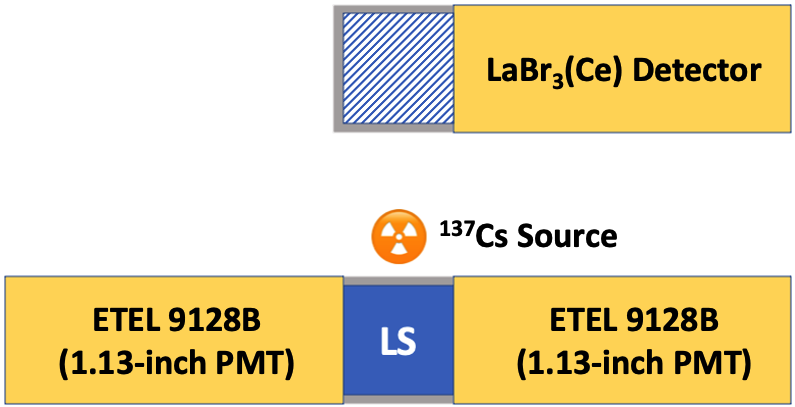}
    \caption{Schematic drawing of the~LY setup. A small~$1\times1$~inch highly reflective~{PTFE} liquid scintillator cell with thin UV-transparent glass windows is filled with the LS sample. Two~PMTs are coupled to the cell. The LS is irradiated by monoenergetic gamma-quanta from a~$^{137}$Cs source. To fix the scattering angle and by that the energy deposition of the gammas in the LS, the setup is triggered in coincidence with a~{LaBr$_3$(Ce)} detector. Only events with a coincident signal in the~185\,keV backscattering pseudo-peak in the crystal are selected in the offline analysis (see also \autoref{fig:BackscatterLYOverviewSpecCorr}).}
    \label{fig:LySetupMZ}
\end{figure}

\begin{figure}[h!]
    \centering
    \includegraphics[width=0.495\textwidth]{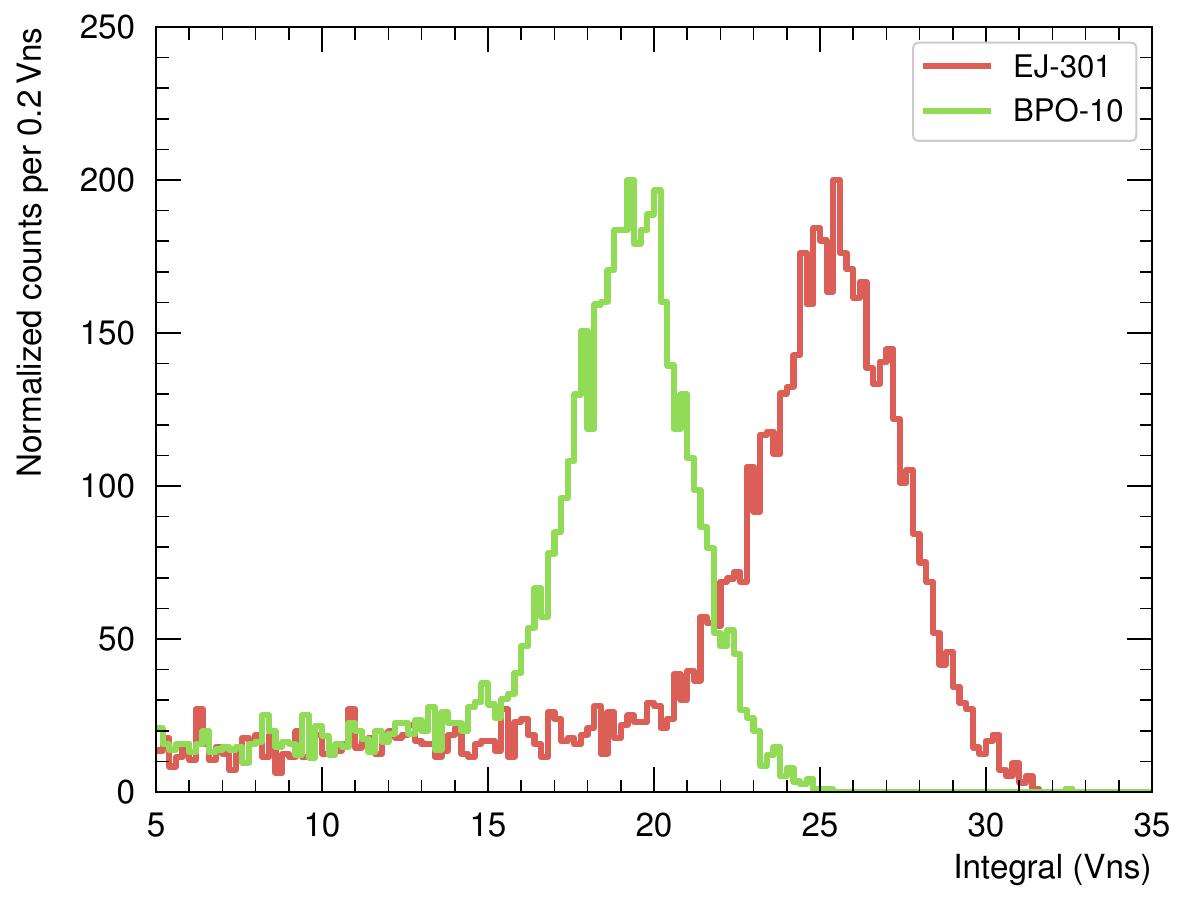}
    \includegraphics[width=0.495\textwidth]{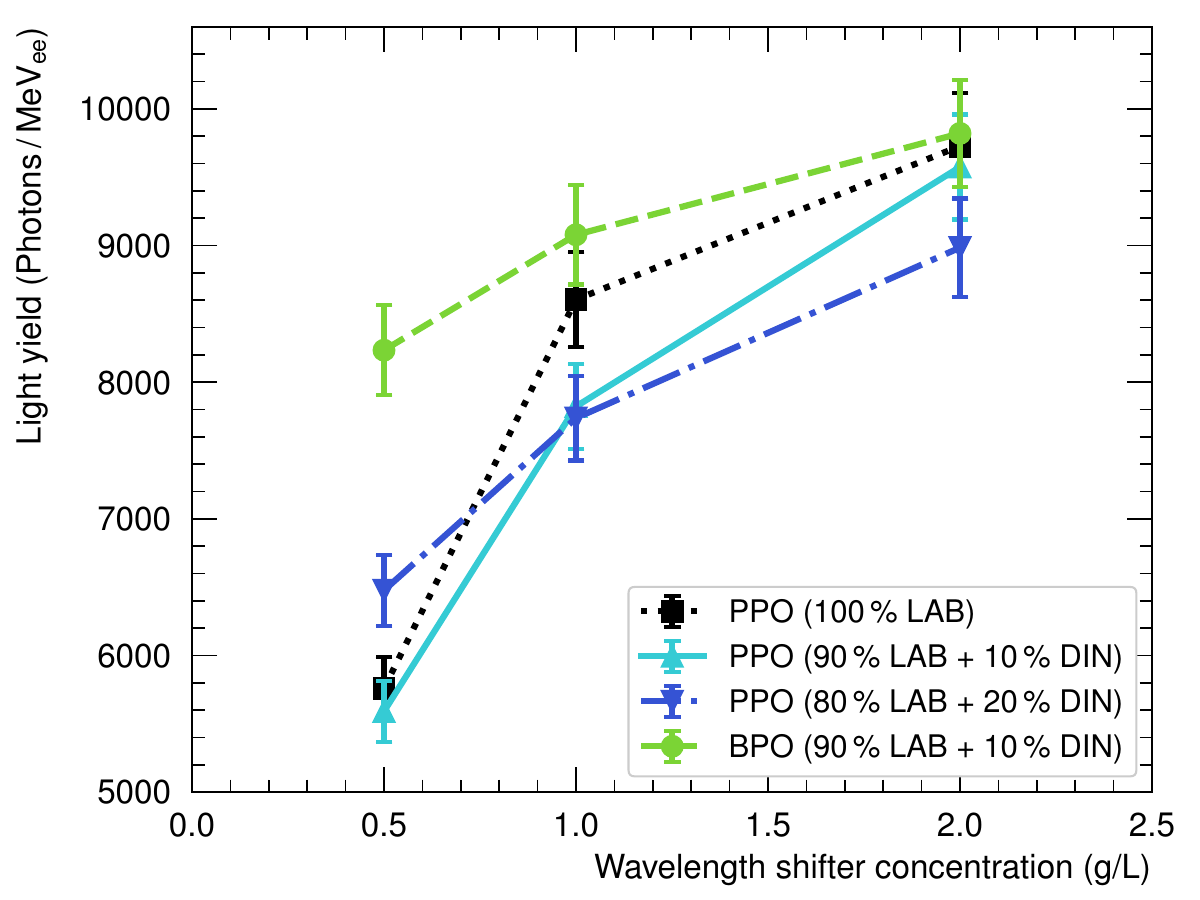}
    \caption{\textbf{Left: }477\,keV peaks from the~180$^{\circ}$ scattering of the gamma radiation from $^{137}$Cs on the LS sample cell filled with the reference~EJ-301 (red) and the~BPO-10 sample (green). \textbf{Right: }Light yield of the samples as a function of fluor concentration. As expected from the known efficient energy transfer from~DIN to~BPO, the sample with this fluor achieves superior light yields already at low admixtures to the solvents (see light green data points).}
    \label{fig:BackscatterLYOverviewSpecCorr}
\end{figure}

\enlargethispage{0.5cm}
The amount of scintillation photons produced for a specific energy deposition by a specific particle species, the LS light yield~(LY), is a crucial parameter greatly affecting the energy resolution and vertex reconstruction capabilities of a scintillation medium. In a two component scintillator (solvent + fluor) the LY drastically increases with increasing fluor loadings from approx.~{1\,g/l\,$-$\,10\,g/l}. Here, internal losses in the solvent (mainly due to impurities) compete with the energy transfer from the solvent to the fluor. At the critical concentration~(typ.~{$<1$\,g/l}) for the solvent-fluor combination, about half the maximum light yield is reached~\cite{BuckYeh, AberleChem}. For scintillation cocktails with effective energy transfers between solvent and fluor the critical concentration is lower, leading to higher light yields with lower amounts of fluor.\\ 
Absolute measurements of the amount of scintillation photons emitted for a given energy deposition in a scintillator liquid are very difficult and prone to mismeasurements. Therefore, most light yields are determined relative to a reference scintillator with a known LY~\cite{Bonhomme}. For the LY studies presented here, this method was also applied with {EJ-301} as a reference. To minimize the systematic uncertainty samples and reference should be measured in the same setup. The LY of {EJ-301} and similar scintillators ({BC-501A}, {NE-213}) is stated by the suppliers consistently to reach~78\% of anthracene, which produces~$\sim$17400\,Photons/MeV~\cite{Bonhomme}. Given those numbers the LY of {EJ-301} can be calculated to be~$\sim$13572\,Photons/MeV.\\ 
To evaluate the light yield of the liquid scintillators, the setup shown in \autoref{fig:LySetupMZ} was used. A~12.9\,ml sample of LS is filled under protective~N$_2$ atmosphere into a highly reflective~PTFE cell with~2\,mm thin~UV-transparent glass windows such that the entire volume is free of gas bubbles. Before and after the filling the sample as saturated with protective inert gas by extensively bubbling it with~N$_2$. To protect the sample, the cell is sealed under moderate over-pressure. For the readout of the scintillation light, two conventional~1.13-inch~PMTs~(ETEL9128B) are coupled to the cell windows. During the measurements the sample is irradiated by mono-energetic~662\,keV gammas from a~$^{137}$Cs source (activity:~$\sim$370\,kBq). The position of the source is thereby fixed with respect to the LS cell. The setup is triggered on a coincidence with a~1.5\,inch~$\times$~1.5~inch~LaBr$_3$(Ce) detector provided by~OST Photonics with~3.05\% resolution at~662\,keV. In the offline analysis only events with a coincident signal in the~185\,keV backscattering pseudo-peak of the crystal detector and thus a~477\,keV energy deposition in the liquid scintillator~(see~\autoref{fig:BackscatterLYOverviewSpecCorr} left) are selected.

\FloatBarrier

By studying the reproducibility of the light yield measured for the reference sample~(EJ-301, red histogram in \autoref{fig:BackscatterLYOverviewSpecCorr} left) the combined systematic uncertainties were estimated. The~EJ-301 sample was measured~10 times on different days each time freshly filled into the previously cleaned empty cell. The variance of these measurements is interpreted as the influence of systematic effects like~HV stability, minor temperature changes in the laboratory, efficiency of the oxygen removal from the LS and other sample preparation and cleaning related differences. By this study the relative systematic uncertainty on the~LY of a given sample can be estimated to a value of~3.9\%, while statistical errors of~$\sim$0.1\% were achieved for the measurements listed in \autoref{LyTabs} and visualized in a comparison plot (see \autoref{fig:BackscatterLYOverviewSpecCorr} right).

\FloatBarrier

\begin{table}[h]
\centering
\begin{tabular}{|c|c|c|c|}
\hline 
Sample Name & Rel. LY in \% & LY [Ph./MeV] & LY Spectr. Corr. [Ph./MeV] \\ 
\hline 
Anthracene & 100 & 17400 & - \\
\hline 
EJ-301 & 78 & 13572 & 13572 \\
\hline 
LAB + 0.5\,g/l PPO & 39.5 $\pm$ 1.6 & 6877 $\pm$ 275 & 5756 $\pm$ 230  \\
\hline 
LAB + 1.0\,g/l PPO & 59.1 $\pm$ 2.4 & 10281 $\pm$ 412 & 8605 $\pm$ 345 \\
\hline 
LAB + 2.0\,g/l PPO & 66.8 $\pm$ 2.7 & 11622 $\pm$ 465 & 9728 $\pm$ 390 \\
\hline 
PPO-05 & 38.4 $\pm$ 1.6 & 6679 $\pm$ 268 & 5590 $\pm$ 225 \\
\hline 
PPO-10 & 53.7 $\pm$ 2.2 & 9345 $\pm$ 374 & 7822 $\pm$ 314\\
\hline 
PPO-20 & 65.7 $\pm$ 2.7 & 11440 $\pm$ 458 & 9575 $\pm$ 384 \\
\hline 
BPO-05 & 53.8 $\pm$ 2.2 & 9367 $\pm$ 375 & 8234 $\pm$ 330 \\
\hline 
BPO-10 & 59.4 $\pm$ 2.4 & 10329 $\pm$ 414 & 9079 $\pm$ 364 \\
\hline 
BPO-20 & 64.2 $\pm$ 2.6 & 11173 $\pm$ 447 & 9821 $\pm$ 393 \\
\hline 
PPO-05-20 & 44.5 $\pm$ 1.8 & 7737 $\pm$ 310 & 6476 $\pm$ 260 \\
\hline 
PPO-10-20 & 53.1 $\pm$ 2.2 & 9244 $\pm$ 370 & 7737 $\pm$ 310 \\
\hline 
PPO-20-20 & 61.7 $\pm$ 2.5 & 10735 $\pm$ 430 & 8985 $\pm$ 360 \\
\hline 
\end{tabular} 
\caption{Listed are the samples, that have been investigated in the~LY setup together with anthracene, whose~LY is~100\% by definition. The commercial~{EJ-301} serves as the reference, while also conventional LAB/PPO mixtures were measured to allow a comparison of the~LY results with other publications~(e.g.~\cite{Bonhomme}). Column~2 depicts the relative light yield compared to anthracene, while the values in coulumn~3 and~4 are converted in absolute values. For the values in the right column the effect of different quantum efficiencies of the~PMTs for the emission spectra of the samples as well as different spectral transmission of the cell windows were corrected.}
\label{LyTabs}
\end{table}

\FloatBarrier


\subsection{Separation of Scintillation and Cherenkov Light}
\label{ChS}

\begin{figure}[ht!]
\centering 
\includegraphics[width=0.50\textwidth]{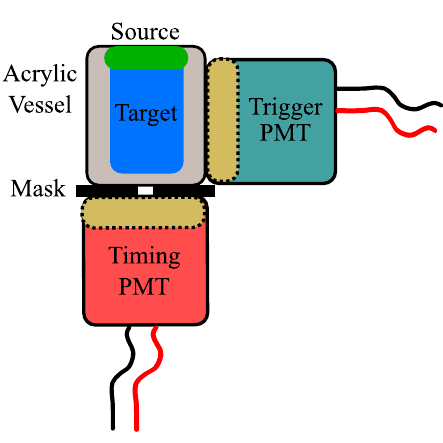}
\caption{The experimental setup to measure the Cherenkov and scintillation separation. The trigger~PMT is a~1-inch~H11934-200 Hamamatsu~PMT with an ultra bialkali photocathode and the timing~PMT is a~1\,inch~{H11934-20} Hamamatsu with an extended red multialkali photocathode. The black mask has a~2\,mm diameter to limit the coincidence rate and maintain primarily single photoelectron detection at the timing~PMT. The target material is water,~PPO-05, or~PPO-10.}
\label{fig:timing_setup}
\end{figure}

To study the separation of scintillation and Cherenkov light emission in the scintillator samples the experimental setup shown in \autoref{fig:timing_setup}, based on the well-established time-correlated single photon counting~(TCSPC) method \cite{Bollinger, Marrodan}, was used. An acrylic vessel~(AV) with a height of 3\,cm and an identical diameter containing a sample volume of~$\sim$7.6\,ml was sealed by a disc-shaped~$^{90}$Sr~$\beta$ source above the LS. Thus, the source irradiated the sample directly (without an additional acrylics particle entrance window) with~$\beta$ radiation. The~$^{90}$Sr~$\beta$ decays with~Q-value of~0.55\,MeV to $^{90}$Y, which subsequently~$\beta$ decays with a half-life of~$\sim$64 hours and a particularly high Q-value of~2.28\,MeV. A 1-inch Hamamatsu H11934-200~PMT (referred to as the trigger~PMT) with a rectangular ultra-bialkali photocathode~(UBA), is optically coupled to the side of the AV using Eljen Technology~EH-550 optical grease. Another~1-inch Hamamatsu~H11934-20~PMT (referred to as the timing~PMT), with an extended red multi-alkali photocathode~(ERMA), is coupled to the acrylic vessel from below. The typical transit-time spread of these~PMTs reaches $\sim$270\,ps~(FWHM) \cite{HamamatsuDataSheet}. An aperture (made from black highly light absorbing material) with a~2\,mm diameter central hole was placed between the timing~PMT and the acrylic vessel in order to limit the amount of light observed by the~PMT. Due to this mask, the timing~PMT observes primarily single photons. 

The signals from both~PMTs are digitized using a~CAEN~V1742 digitizer \cite{CaenV1742} over a 1\,V dynamic range, sampling at~5\,GHz for~1024 samples. The signal from the trigger~PMT is used to trigger the data acquisition. The data is readout via USB using custom~DAQ software \cite{CHESS}. The waveforms for both~PMTs are processed offline. In a post-trigger window, timing~PMT pulses are identified by looking for samples that cross a threshold of~2.5\,mV. For events where a~PMT pulse is identified, the time is calculated by applying a software-based constant fraction discriminator to the pulse. For the trigger~PMT, the time is assigned by identifying the~3\,mV crossing time and used to provide a time-zero. For both~PMTs, a linear interpolation between the sampling point of the trace was applied to further refine the time measurement. The time difference between the registration of the single photoelectron at the timing~PMT and the time-zero from the trigger~PMT is referred to as~$\Delta t$. It contains information about the the emission timing of the photon and hence Cherenkov and scintillation separation. This algorithmic implementation of the~TCSPC closely follows the ones used in previous Cherenkov and scintillation separation studies~\cite{LAPPDseparation}.

\begin{figure}[h]
\centering
\includegraphics[width=0.6\textwidth]{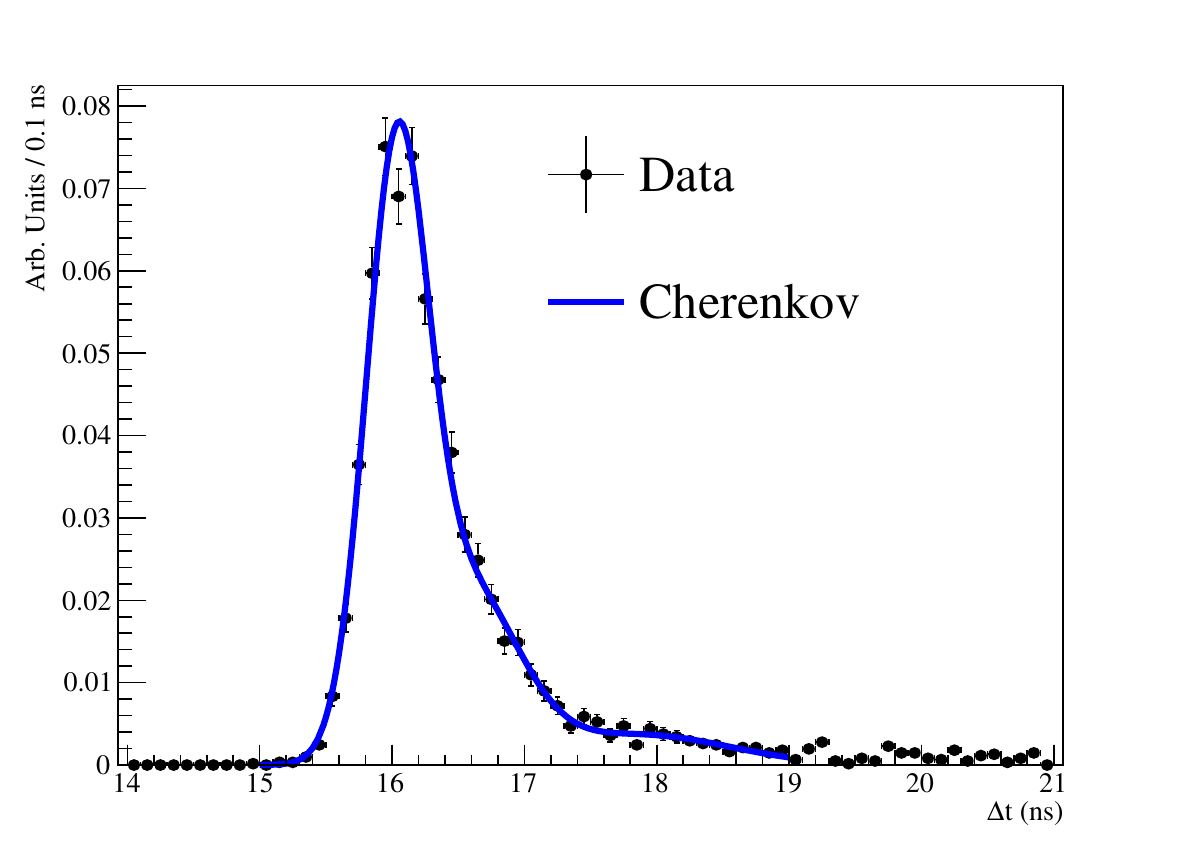}
\label{fig:water_berkeley}
\caption{The measured timing for~$\beta$ excitation of a water sample, with the corresponding fit consisting of the sum of three Gaussian distributions.}
\end{figure}

In order to extract the instrumental response of the setup, which is later used for modeling the light emission measurements, a dataset is recorded with water as target medium in the~AV. This provides a pure Cherenkov light dataset to extract the overall response of the system~($F(t)$) to a prompt signal.~\autoref{fig:water_berkeley} shows the so measured data with a fit that includes the sum of three Gaussian distributions. This empirically determined response contains a primary peak as well as a secondary shoulder mainly caused by single photoelectron late pulsing of the timing~PMT. The width and relative fractions of those distributions are held fixed in the light emission model later ultimately applied to the data of the LS.

\enlargethispage{1cm}

\begin{figure}[h]
    \centering
    \includegraphics[width=0.45\textwidth]{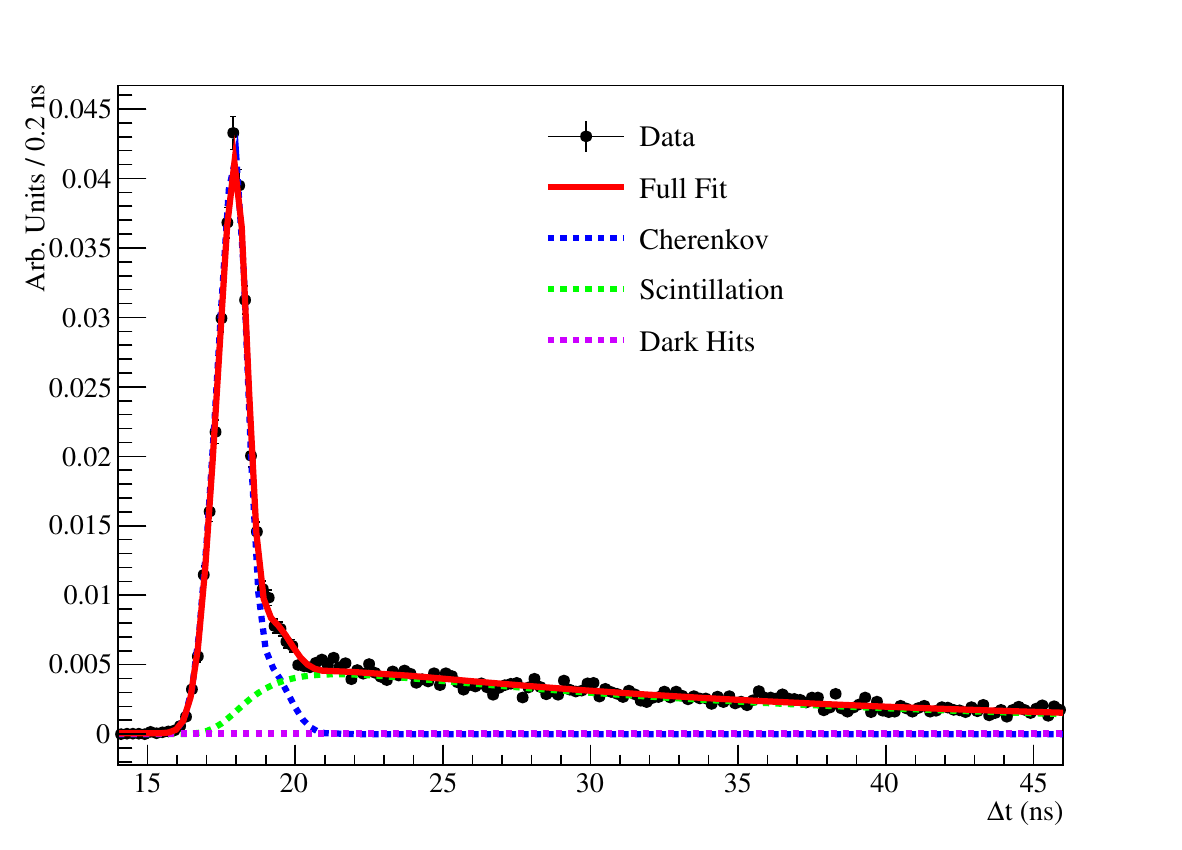}
    \includegraphics[width=0.45\textwidth]{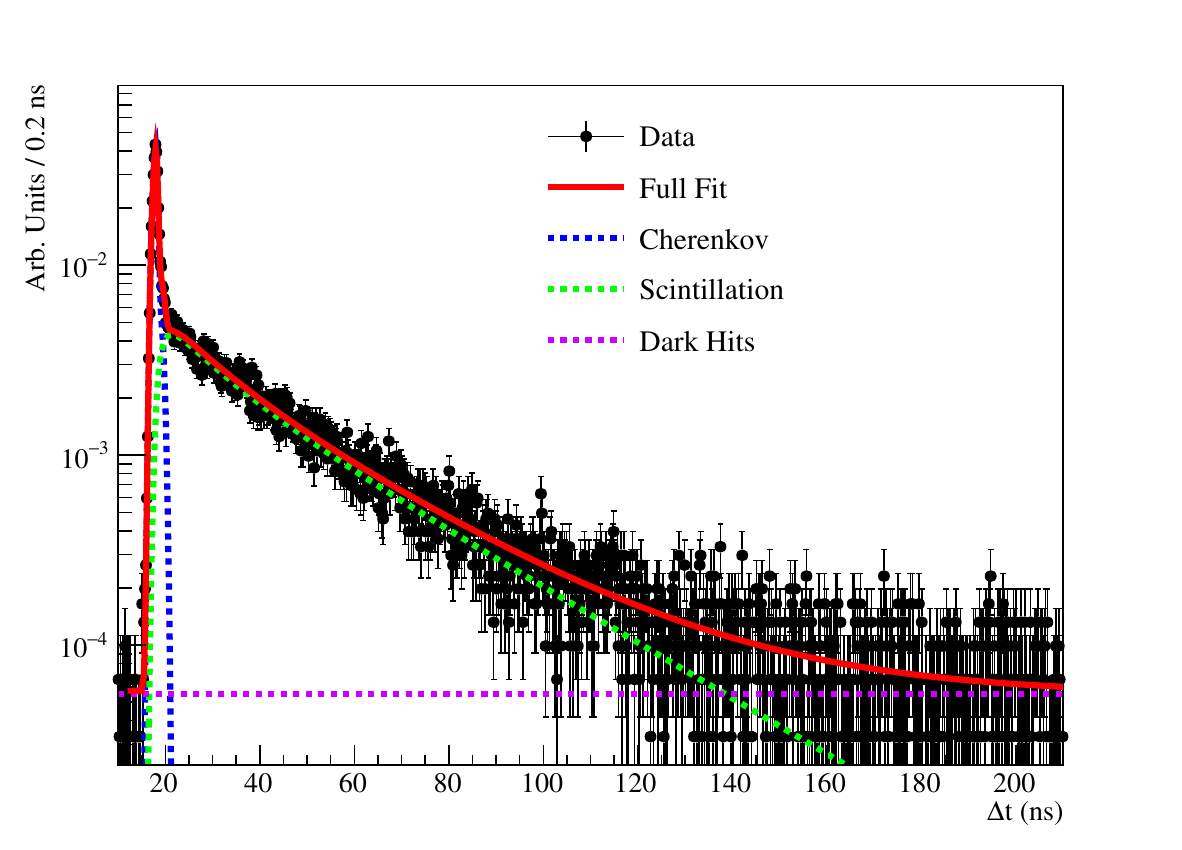}
    \includegraphics[width=0.45\textwidth]{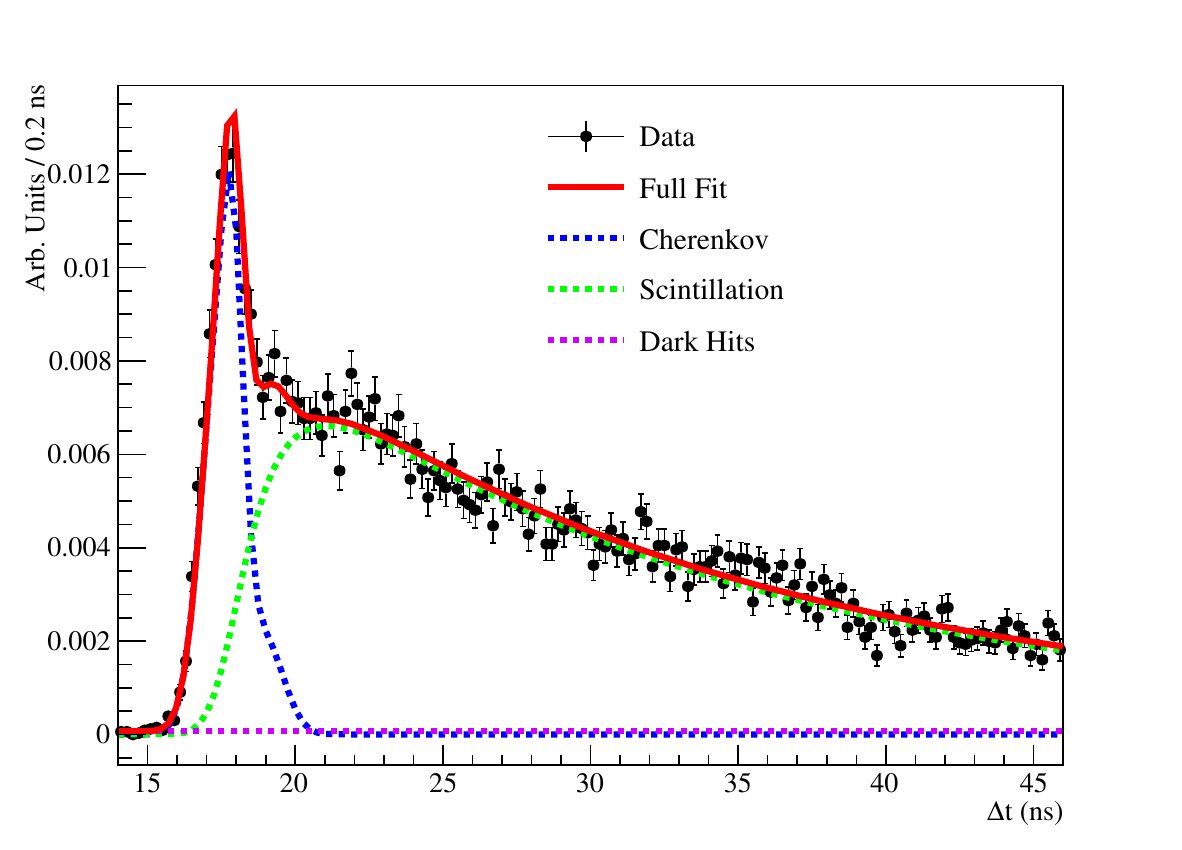}
    \includegraphics[width=0.45\textwidth]{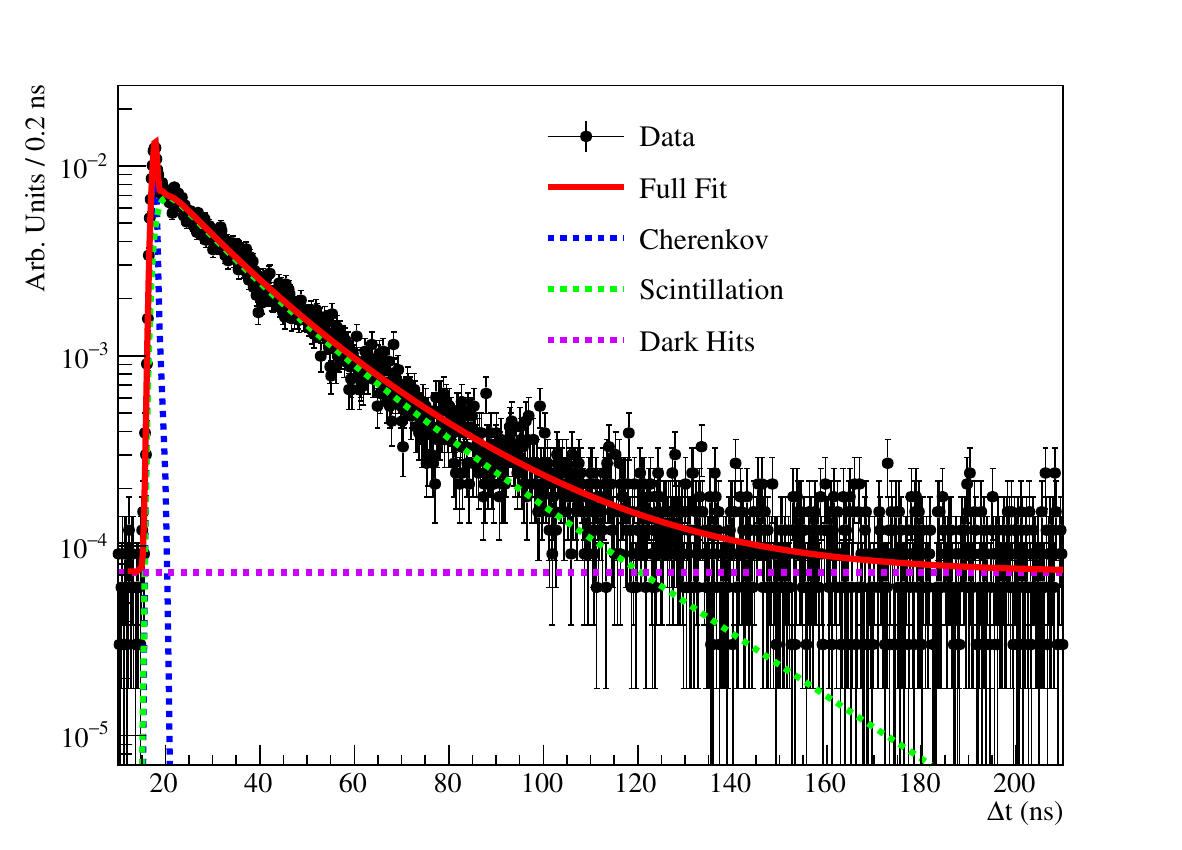}
    \caption{The measured emission timing for~$\beta$ excitation of samples~(top)~PPO-05 and (bottom)~PPO-10. The measurements are shown in linear scale around the Cherenkov peak~(left) and in log scale to highlight the scintillation tail~(right).}
    \label{fig:time_profiles_berkeley}
\end{figure}

For this publication, the~PPO-05 and~PPO-10 samples were examined in detail for Cherenkov and scintillation separation.~\autoref{fig:time_profiles_berkeley} shows the data for the corresponding measurements. Notably, there is a clearly identifiable Cherenkov component for both samples visible as a dominant peak in the beginning of the light emission time profile. As the light yield of~PPO-05 is lower than~PPO-10 and also the effective scintillation light emission is slower, the Cherenkov peak is here more prominent. To quantify this, the purity of the Cherenkov light in this peak is defined as: 
\begin{equation}\label{eq:purity}
    P  = \frac{\int_{-\infty}^{t_{f}} C(t) dt}{\int_{-\infty}^{t_{f}} (C(t) + S(t)) dt},
\end{equation}
where~$C(t)$ and~$S(t)$ denote the fitted Cherenkov and scintillation components respectively. In the integral,~$t_{f}$ is optimized by maximizing:
\begin{equation}\label{eq:r}
    R(t_{f}) = P(t_{f}) \times \int_{-\infty}^{t_{f}} C(t) dt, 
\end{equation}
which is equivalent to optimizing the standard signal-to-background metric~$C/\sqrt{C + S}$. This method ensures the window selection does not optimize only purity, which would tend to select a very narrow and early-time window, but also the total number of Cherenkov photons within this window.
\enlargethispage{1.25cm}
In addition to calculating a Cherenkov purity, the data is fit with a model for the scintillation light emission according to:
\begin{equation}\label{eq:scint_time_profile}
    S(t)
        = \sum_{i=1}^{2}
            A_{i} \Theta(t) \frac{e^{-t/\tau_{i}} - e^{-t/\tau_{r}}}{\tau_{i} - \tau_{r}}
\end{equation}
where
\begin{equation}\label{eq:norm}
    \sum_{i=1}^{2} A_{i} = 1,
\end{equation}
and~$\tau_{i}$ are the scintillation emission decay-time constants, while~$A_{i}$ denote the associated normalization factors, and~$\tau_{r}$ the scintillation rise-time. The functional form for the scintillation emission is convoluted with the detector response,~$F(t)$, derived from the fitted water data. Due to the short acquisition window from the~CAEN~V1742 digitizer of~200\,ns at 5\,GS/s long and less prominent decay components of the scintillation time profile are not visible (see \autoref{fig:time_profiles_berkeley}) and thus both data sets can be modeled with two exponential decays.  

\FloatBarrier

The parameters obtained from the fits as well as the derived Cherenkov purities are given in \autoref{tab:time_profile_berkeley}. The measured rise-times exceed~1\,ns for both samples. The primary difference in the time-profile for each sample can be identified in the fraction of light in the fastest emission component,~$A_{1}$, which is twice as large for~PPO-10 as for~PPO-05. Notably, both samples (containing the conventional fluor~PPO in concentrations of~0.5\,g/l and~1.0\,g/l respectively) are by far slower than the scintillators used in currently operating liquid scintillator detectors, such as~SNO+, where the scintillator has a~$\tau_{1}\sim$~5\,ns~\cite{SNOpScint}. 

\begin{table}[h]
    \centering
    \begin{tabular}{|c|c|c|c|c|c|c|}
         \hline 
         Sample & $A_{1}$ & $A_{2}$ & $\tau_{r}$ (ns) & $\tau_{1}$ (ns) & $\tau_{2}$ (ns) & $P$ (\%) \\ \hline 
         PPO-05 & 0.21 $\pm$ 0.06 & 0.79 $\pm$ 0.06 & 1.40 $\pm$ 0.40 & 11.7 $\pm$ 3.2 & 29.1 $\pm$ 3.5 & 91.0 \\ \hline 
         PPO-10 & 0.44 $\pm$ 0.07 & 0.56 $\pm$ 0.07 & 1.20 $\pm$ 0.43 & 13.0 $\pm$ 3.4 & 26.3 $\pm$ 4.6 & 80.3 \\ \hline 
    \end{tabular}
    \caption{The measured scintillation emission timing parameters for~$\beta$ excitation of~PPO-05 and~PPO-10.}
    \label{tab:time_profile_berkeley}
\end{table}

\FloatBarrier

\subsection{Scintillation Time Profiles and Pulse Shape Discrimination Capabilities}

One of the most powerful background suppression techniques in LS detectors is pulse shape discrimination. Pulses, that~PMTs record, are influenced by the hardware itself as well as more fundamentally are built up by the intrinsic fluorescence time profile of the~LS. This profile depends on the one hand on the scintillator's composition and on the other on the differential energy loss~{d$E/$d$x$} of the particles traversing the LS (for theoretical descriptions and further empirical details see~\cite{Förster, Birks, Horrocks}). The~PSD performance of a~LS is closely related to differences in the time spectrum of the photon emission for different particle species. To evaluate this potential also in the new LS media with slow light emission, a detailed study of the scintillation time profiles for neutron (proton recoils in the LS) and gamma radiation (electron recoils) signals was performed. 

\subsubsection{Experimental Setup}
\label{ExpSetupSubs}
\begin{figure}[h!]
    \centering
    \includegraphics[width=1.0\textwidth]{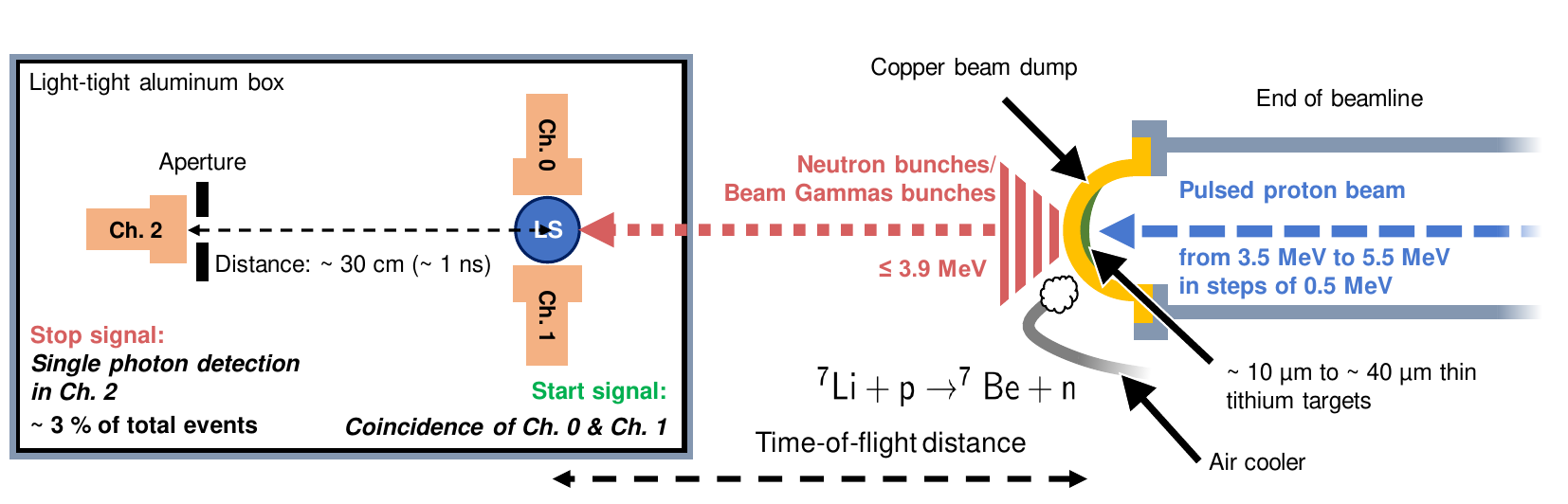}
    \caption{Illustration of the scintillation time profile experiment at the end of the beamline of the~CN Van de Graaff accelerator at~INFN Legnaro. The experimental setup exploits the~TCSPC technique. The start signal of the time measurement is gained from the coincident pulsing of two PMTs~(Ch.~0 and Ch.~1) directly on the LS vessel. A stop signal is generated, when a~PMT~(Ch.~2) in a distance of~60\,cm placed behind an aperture detects a single photon. Irradiation with quasi-monoenergetic neutrons is possible by bombarding thin metallic~Li targets with protons. A copper backing cooled by forced air acts as beam stopper.}
    \label{fig:FluorSetup}
\end{figure}

The setup shown in~\autoref{fig:FluorSetup} developed for detailed studies of the~JUNO-LS in \cite{HansPhD, RaphaelPhD} was reused. To allow neutron and gamma irradiation, it was placed at the end of the~0$^{\circ}$ beamline of the~CN accelerator at~INFN Legnaro. Here, pulsed proton beams with energies from~3.5\,MeV up to ~5.5\,MeV~($\sigma_{E}\sim$~3\,keV) can be guided onto a thin metallic lithium target~(typ.~5-20\,$\upmu$m). In the nuclear reaction 

\begin{equation}
\text{p} +~^{7}\text{Li} \longrightarrow~^{7}\text{Be} + \text{n}
\end{equation}

quasi-monoenergetic neutrons~(QMN) adjustable between~3.86\,MeV down to~1.86\,MeV as well as beam correlated gammas can be created. The detector setup is placed in a distance of~1.5\,m from the lithium target. The bunch widths of typ. well below~1\,ns and a repetition rate of~600\,kHz allow time-of-flight~(ToF) discrimination of neutrons and gammas.\\
The experiment determines the probability density function of the scintillation light emission similarly to the setup shown in~\ref{ChS} by~TCSPC. The detector setup for the fluorescence time profile measurement is placed onto optical benches in a darkbox~(wall thickness~1\,mm) made of aluminum that is also acting as a faraday cage. A spherical borosilicate glass vessel with an outer diameter of~$\sim$72\,mm and wall thickness below~1\,mm contains a sample of~$\sim$150\,ml~LS. To prevent oxidation of the LS, the remaining volume in the sphere is filled with a protective nitrogen atmosphere with an overpressure of some millibar. The vessel is enclosed by two gas-tight stopcocks with~PTFE plugs. The sphere with the~LS is placed between two~ETEL~9821B~PMTs with a~68\,mm photocathode~\mbox{\cite{HansPhD, StockProceed}}. 

The coincident pulsing of these tubes provides the start signal for the time measurement. The time of the energy deposition of an incident particle in the~LS is determined by calculating the mean of the pulses' onsets by means of a constant fraction algorithm. A third~PMT of the same type is placed in a distance of approximately~60\,cm from the sphere and behind an adjustable aperture. By varying the aperture's opening diameter, the single photon detection probability at the distant~PMT can be fine adjusted to~$\sim3\%$. A single photon hit of this tube provides the stop signal for the time measurement, which is also extracted offline from the~PMT's waveform by the constant fraction algorithm~\cite{HansPhD}. 

The PMT traces with a total length of~1000\,ns were digitized by a high-performance flash~ADC (Agilent Acqiris,~10\,bit,~2\,GS/s). The trigger logic comprises a flexible and fast~NIM-based system, which can be adjusted for efficiently working with either a radioactive sources such as~$^{137}$Cs or the particle accelerator including also a coincidence with the beam bunches. The system is adjusted such that time differences of the start and stop signals up to~650\,ns can be recorded. To avoid pile-up of events or multi-particle hits of the detector in one bunch, the analog trigger rate of the setup was kept well below~2\,kHz~(coincidence of~Ch.~0 and~Ch.~1 with the beam bunch) by reducing the beam current hitting the Li-target below 20\,pnA. The instrumental response function~(IRF) was studied with a picosecond pulsed diode laser system provided by~NKT Photonics with a typical pulse width below~$\sigma=9$\,ps~\cite{HansPhD, RaphaelPhD, StockProceed}.


\FloatBarrier

\subsubsection{Data Modeling}
\label{DataModeling}
\begin{figure}[h!]
    \centering
    \includegraphics[width=1.0\textwidth]{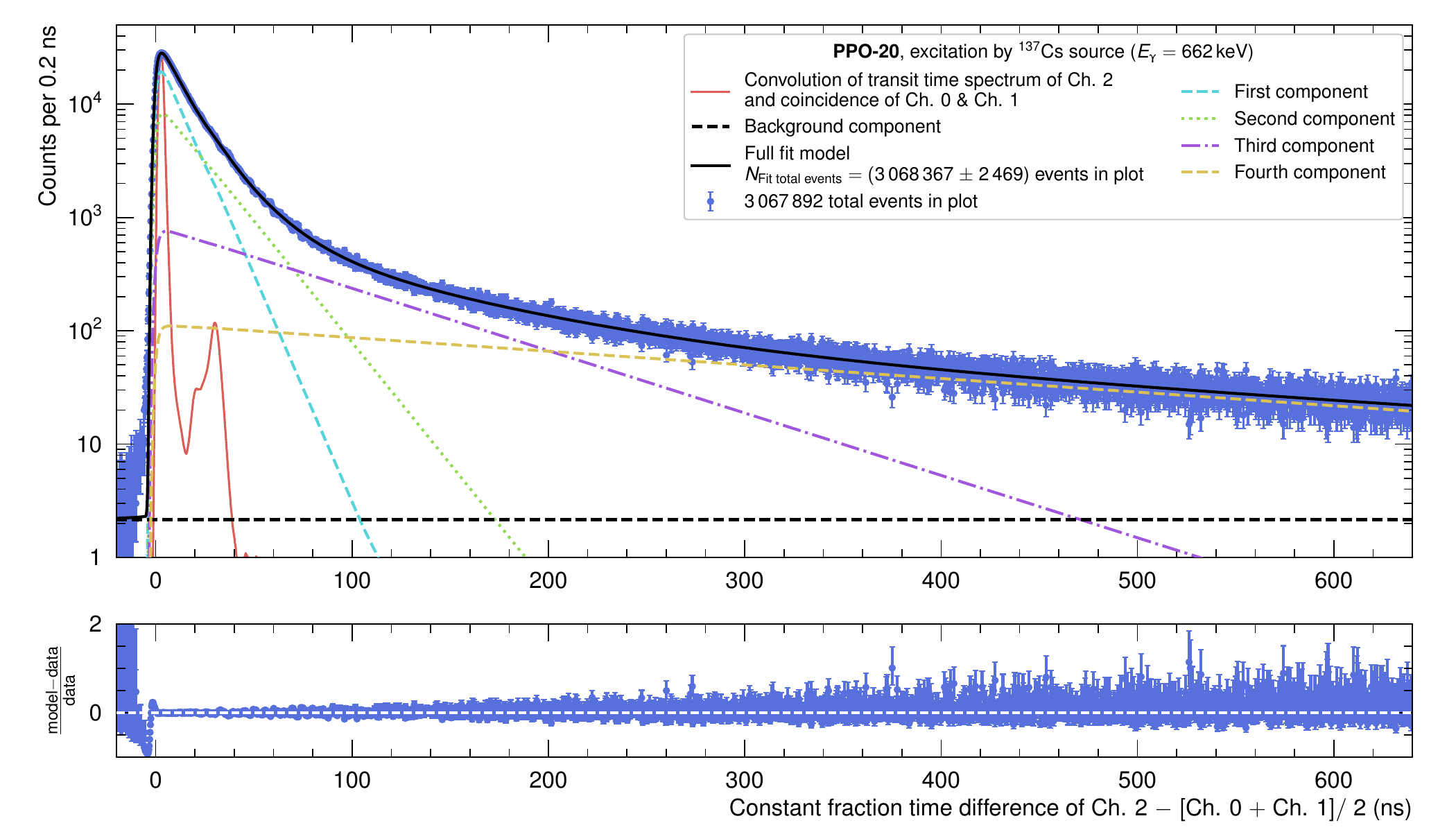}
    \caption{Scintillation time profile of the~PPO-20 sample for the irradiation with gammas from a~$^{137}$Cs source. The~IRF and the components of the scintillation decay model are also depicted separately. The results obtained from the fit~(solid black line) can be found in \autoref{PSDTab2}.}
    \label{fig:DecaySpecCompare}
\end{figure}

The scintillation time profile is modeled in a similar way as in~\ref{ChS} but including also two longer decay components present in the long tail of the light emission spectrum~(see~\autoref{fig:DecaySpecCompare}). Therefore, the model can be depicted as 

\begin{equation}\label{eq:scint_time_profileStock1}
    S(t)
        = \sum_{i=1}^{4}
            A_{i} \Theta(t) \frac{e^{-t/\tau_{i}} - e^{-t/\tau_{r}}}{\tau_{i} - \tau_{r}}
\end{equation}
where
\begin{equation}\label{eq:scint_time_profileStock2}
  \sum_{i=1}^{4}
            A_{i}  = 1.
\end{equation}

Following~\cite{Lifetime}, the intensity-weighted average lifetime~$\tau_{\text{life}}$ of the scintillation light emission can be calculated by 
\begin{equation}\label{eq:EffectLifeStock}
    \tau_{\text{life}}
        = \sum_{i=1}^{4}
            A_{i} \tau_{i} + \tau_r.
\end{equation}
Since here a different data parameterization than in~\cite{Lifetime} is applied, the scintillation rise time can be included.\\
\enlargethispage{1.0cm}  
As mentioned in~\ref{ExpSetupSubs} the~IRF is measured by means of a picosecond laser. The time response of the setup is governed by the transit time spectrum for single photon hits of the distant PMT~(Ch.~2). To also account for the smearing of the time profile start signal, which is partially caused by the~LS sample but mainly by the time jitter of the used~PMTs, the in-situ gained data for this coincidence is convoluted. The resulting~IRF is drawn as the red solid line in \autoref{fig:DecaySpecCompare} and shows next to the nearly Gaussian main peak the expected distinct late pulsing population. The fit model is thereby the numerical binwise convolution of the~IRF with the four exponential decays including the rise time and an additional constant background component (see dashed black line in \autoref{fig:DecaySpecCompare}). The results of all fits applied to the data for both, neutron and gamma interactions, are given in \autoref{PSDTab1} and \autoref{PSDTab2}.     

\FloatBarrier

\subsubsection{Evaluation of the Pulse Shape Discrimination Capability}

\begin{figure}[h!]
    \centering
    \includegraphics[width=0.495\textwidth]{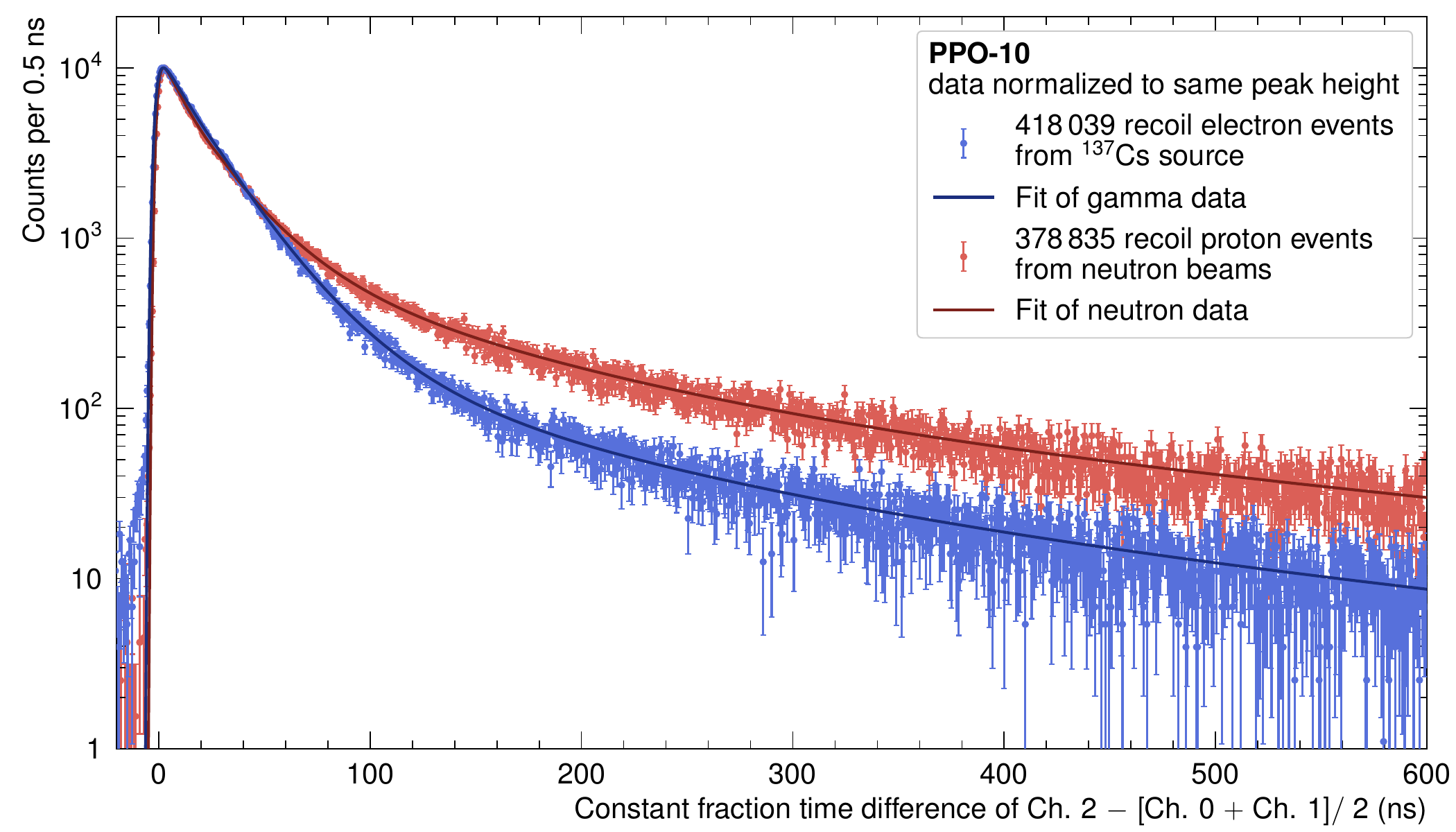}
    \includegraphics[width=0.495\textwidth]{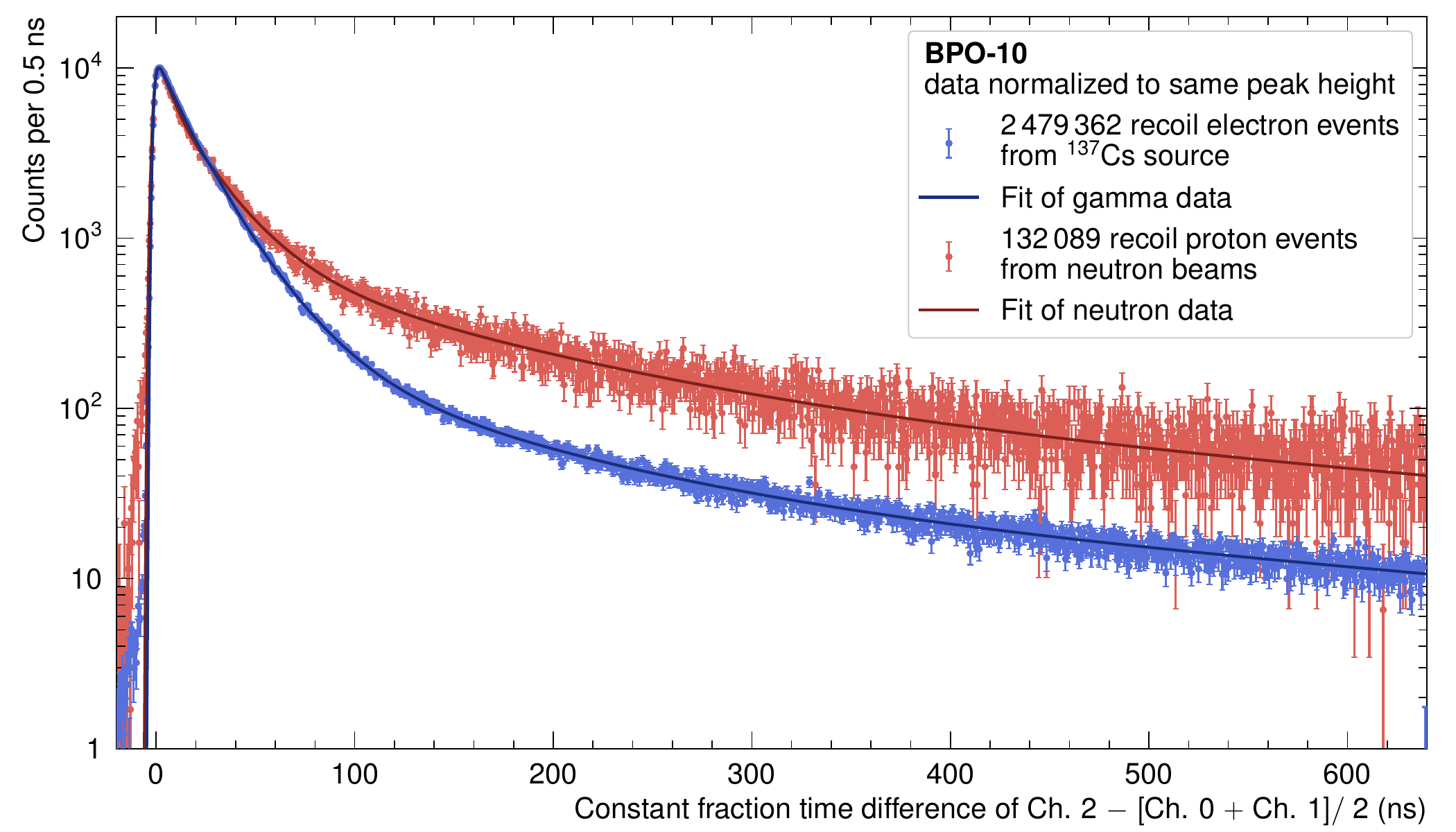}
    \includegraphics[width=0.495\textwidth]{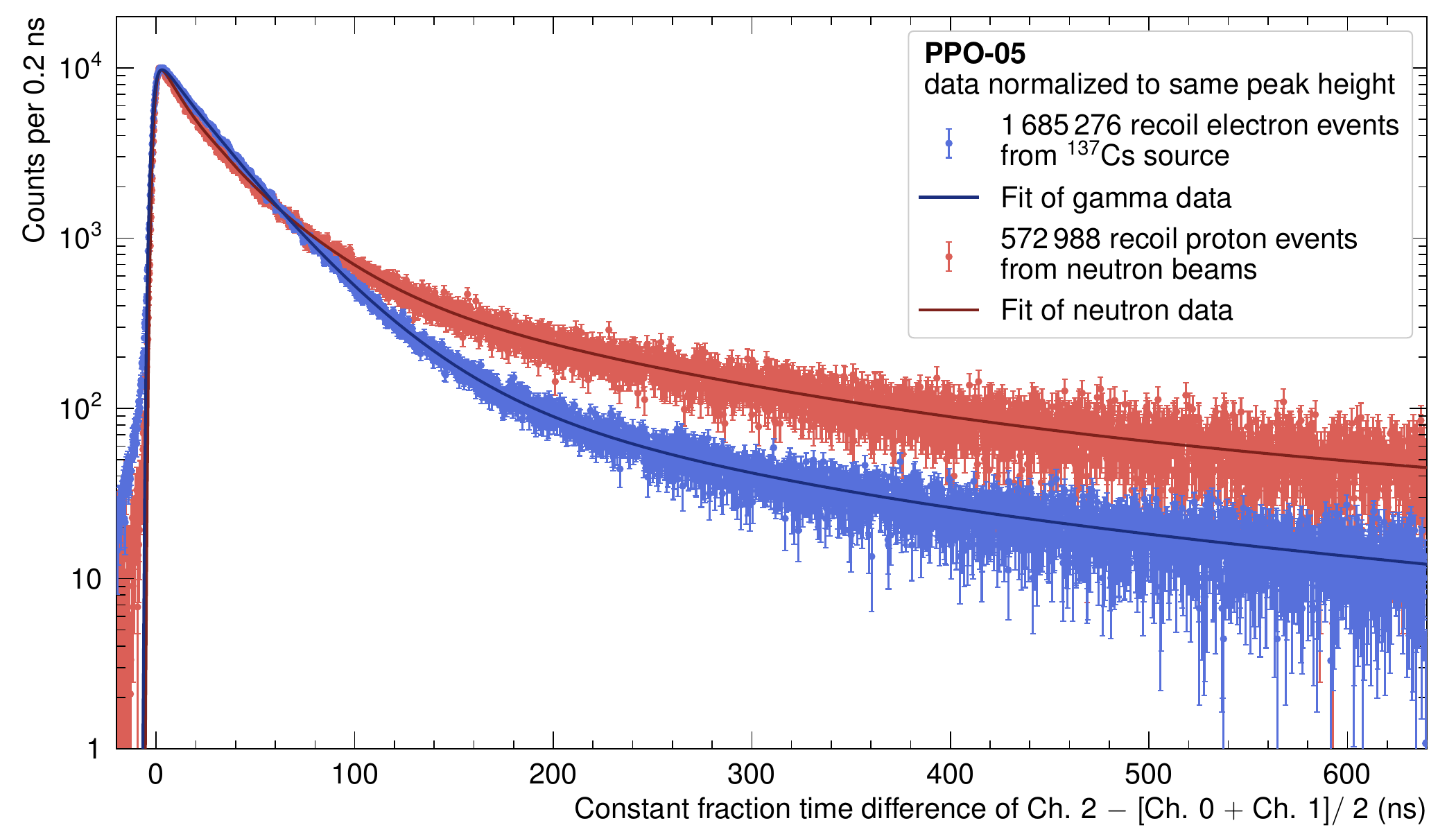}
    \includegraphics[width=0.495\textwidth]{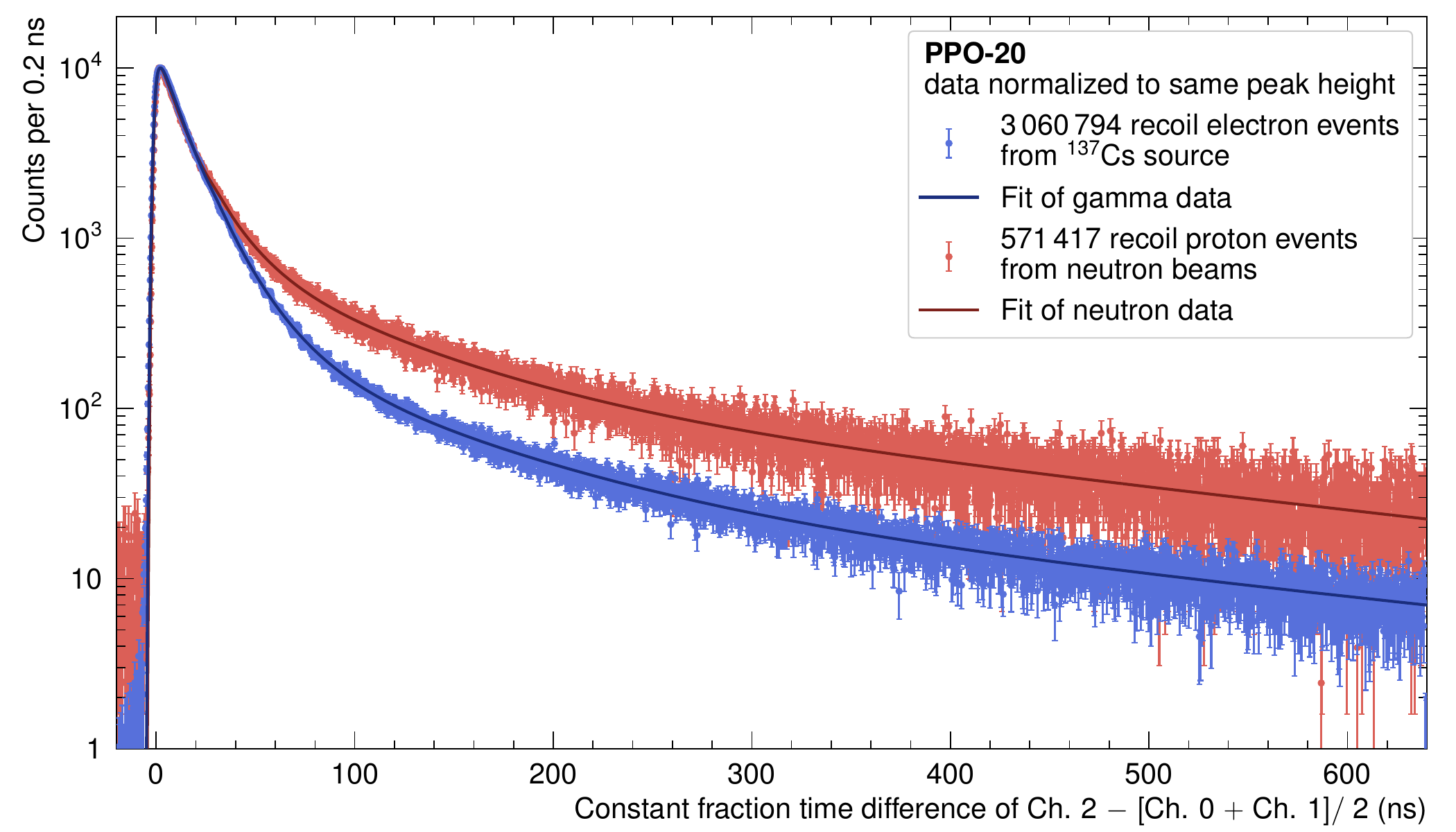}
    \caption{Electron recoil (blue) and proton recoil (red) time profiles of the liquid scintillators PPO-10~(top left), BPO-10~(top right), PPO-05~(bottom left) and PPO-20~(bottom right). The corresponding fits according to the model described in \ref{DataModeling} are drawn as black solid lines}
    \label{fig:time_profiles_tum_lnl}
\end{figure}

To evaluate the~PSD capability, the~QMN beam was used to irradiate the~LS candidates. As the available beamtime was limited the samples~PPO-05,~PPO-10,~PPO-20 and~BPO-10 were selected for the study as they all contain an admixture of~10\% DIN. Given the time structure of the beam, neutrons and~$\gamma$s were distinguished by~ToF, realizing a purity of the neutron sample well above~99.8\% in all measurements. As the yield of the~$\gamma$-flash from the protons bombarding the Li-target is low, the statistics for the electron recoil profiles was significantly enhanced by the use of a~$^{137}$Cs source with~$A=370$\,kBq. By doing so, large statistics of~$O=3\times10^5-3\times10^6$ electron or proton recoil events (after data quality cuts) containing a single photons hit in~Ch.~2 was gained. \autoref{fig:time_profiles_tum_lnl} shows a comparison of neutron and gamma time profiles for all samples. The depicted time profiles up to~600\,ns after the excitation by an incident particle where fitted according to the procedure described in~\ref{DataModeling}. The corresponding results for electron and proton recoils are listed for~PPO-10 and~BPO-10 in \autoref{PSDTab1} as well as for~PPO-05 and~PPO-20 in \autoref{PSDTab2}.

To evaluate the pulse shape discrimination capabilities of different liquid scintillator mixtures the tail-to-total method, which compares the integrated scintillation tail with the total light emission from the beginning to the end of a scintillation event, is applied. Therefore, the time profile models~(probability density functions), derived from the corresponding fits, are shifted such that their maxima occur at the same position in time. The integration windows for the tail ranging from different start times to~500\,ns after the peak of the time profile. The difference of the areas below the curves (neutron area minus gamma area), is the tail-to-total difference~$\Delta\mu$. The start time of the tail integration after the peak is thereby optimized such, that~$\Delta\mu$ reaches its maximal value.


\begin{table}[h!]
\centering
\begin{tabular}{|c|l|l|l|l|}
\hline 
Sample & \multicolumn{2}{|c|}{PPO-10} & \multicolumn{2}{|c|}{BPO-10}\\
\hline 
Source & e$^-$ & p & e$^-$ & p \\
\hline
$A_1$\,(\%) & $37.76 \pm 8.11 \, ^{+15.59} _{-4.43}$ & $16.48 \pm	4.58 \, ^{+7.72}$ & $25.77 \pm 2.16 \,^{+4.96}$ & $13.08 \pm 2.27 \,_{-1.96}$ \\ 
$A_2$\,(\%) & $49.08 \pm 7.67 \, _{-14.88}$ & $46.71 \pm 3.48 \, _{-5.66}$ & $57.41 \pm 1.99 \,_{-4.54}$ & $43.09 \pm 2.21$ \\ 
$A_3$\,(\%) & $8.02 \pm	1.27$ & $20.97 \pm 1.53$ & $9.34 \pm 0.36$ & $22.74 \pm 2.40$ \\ 
$A_4$\,(\%) & $5.14 \pm	1.46 \, ^{+23.61}$ & $15.84 \pm 2.15$ & $7.48 \pm 0.34 \,_{-0.18}$ & $21.09 \pm 2.88 \,^{+1.92}$ \\ 
\hline
$\tau_r$\,(ns) & $1.60 \pm 0.02 \, ^{+0.07} _{-0.39}$ & $1.19 \pm 0.02 \, ^{+0.30} _{-0.39}
$ & $0.94 \pm 0.01 \,^{+0.18}_{-0.39}$ & $1.24 \pm 0.13 \,^{+0.52}_{-0.39}$ \\
\hline 
$\tau_1$\,(ns) & $16.11 \pm	0.86 \, ^{+1.71}$ & $10.14 \pm	1.11 \, ^{+1.82}$ & $11.41 \pm 0.32 \,^{+0.76}$ & $6.50 \pm 0.92 \,^{+0.68} _{-1.37}$ \\ 
$\tau_2$\,(ns) & $27.25 \pm	1.77 \, ^{+4.11}$ & $24.07 \pm	1.20 \, ^{+2.71} _{-1.05}$ & $22.79 \pm 0.46 \,^{+0.99}$ & $22.17 \pm 1.51$ \\ 
$\tau_3$\,(ns) & $99.23 \pm	20.81 \, ^{+6.97}$ & $90.57 \pm	13.56 \, _{-6.43}$ & $91.46 \pm 5.7 \,^{+7.08}$ & $103.52 \pm 18.66$ \\ 
$\tau_4$\,(ns) &  $347.57 \pm 96.98$ & $373.13 \pm	63.10 \, _{-12.64}$ & $455.53 \pm 38.51 \,^{+55.86}$ & $464.06 \pm 135.01$ \\ 
\hline

$\tau_{\text{life}}$ & $46.89 \pm 7.85 \,^{+82.11}_{-4.14}$ & $92.19 \pm 13.25 \, ^{+1.55}_{-2.84}$ & $59.58  \pm 3.37 \,^{+4.32}_{-1.36} $ & $133.06 \pm 31.84 \,^{+8.89}_{-0.45} $ \\ 
\hline
\end{tabular} 
\caption{Parameters of the electron and proton recoil time profile models for the~{PPO-10} and~{BPO-10} mixtures. Provided values include statistical and asymmetric systematic uncertainties. Although these results used a different analytical form for the electron time profile, in order to have sensitivity to the late-light tail, the obtained values are in good agreement with those presented in \autoref{tab:time_profile_berkeley} for the PPO-10 sample.}
\label{PSDTab1}
\end{table}
\FloatBarrier
\enlargethispage{1.5cm}

\begin{table}[h!]
\centering
\begin{tabular}{|c|l|l|l|l|}
\hline 
Sample & \multicolumn{2}{|c|}{PPO-05} & \multicolumn{2}{|c|}{PPO-20}\\
\hline 
Source & e$^-$ & p & e$^-$ & p \\
\hline
$A_1$\,(\%) & $18.66 \pm 0.20 \,_{-8.06}$ & $14.20	\pm 1.46 \,^{+2.21}$ & $48.45 \pm 2.86 \,_{-5.45}$ & $24.90 \pm 5.32 \,^{+8.87} _{-1.68}$ \\ 
$A_2$\,(\%) & $68.03 \pm 0.26 \,^{+5.06}$ & $48.40 \pm 1.44$ & $34.24 \pm 2.61 \,^{+5.15}$ & $33.75 \pm 4.66 \,^{+0.63} _{-6.83}$ \\  
$A_3$\,(\%) & $7.64 \pm	0.09 \,^{+2.94}$ & $21.46 \pm 1.61$ & $10.62 \pm 0.49$ & $22.13 \pm 1.73$ \\ 
$A_4$\,(\%) & $5.67 \pm	0.06 \,_{-0.38}$ & $15.94 \pm 0.87$ & $6.69 \pm 0.36$ & $19.22 \pm 1.47$ 
\\ 
\hline
$\tau_r$\,(ns) & $1.94 \pm 0.02 \,^{+0.42} _{-0.42}$ & $1.37 \pm 0.02 \,^{+0.43} _{-0.43}$ & $1.46 \pm 0.02 \,^{+0.06}_{-0.19}$ & $1.07 \pm 0.04 \,^{+0.28} _{-0.29}$ \\
\hline
$\tau_1$\,(ns) & $17.79 \pm	0.18 \,^{+0.08}	_{-4.64}$ & $11.53 \pm 0.63 \,^{+1.03} _{-0.16}$ & $10.71 \pm 0.20 \,_{-0.45}$ & $8.83 \pm 0.66 \,^{+1.10}	_{-0.72}$ \\ 
$\tau_2$\,(ns) & $32.40 \pm	0.33 \,_{-2.15}$ & $32.24 \pm 1.21$ & $20.21 \pm 0.82 \,_{-0.95}$ & $17.71 \pm 1.52 \,^{+3.46}$ \\ 
$\tau_3$\,(ns) & $104.68 \pm 1.05 \,^{+5.82} _{-17.88}$ & $135.00 \pm 19.71$ & $78.92 \pm 3.86$ & $67.97 \pm 3.87 \,^{+3.60}	$ \\ 
$\tau_4$\,(ns) & $436.13 \pm 7.52 \,^{+55.57} _{-17.70}$ & $732.53 \pm 361.00$ & $361.22 \pm 20.13$ & $331.59 \pm	15.56 \,^{+13.27}$ \\ 
\hline
$\tau_{\text{life}}$ & $60.03 \pm 0.56 \,^{+4.74}_{-3.27}$ & $164.37 \pm 58.11 \,^{+0.52}_{-0.43}$ & $46.09 \pm 2.07 \,^{+1.05}_{-0.73}$ & $88.03 \pm 6.00 \, ^{+3.05}_{-1.27}$ \\ 
\hline
\end{tabular} 
\caption{Parameters of the electron and proton recoil time profile models for the~{PPO-05} and~{PPO-20} mixtures. Provided values include statistical and asymmetric systematic uncertainties.Although these results used a different analytical form for the electron time profile, in order to have sensitivity to the late-light tail, the obtained values are in good agreement with those presented in \autoref{tab:time_profile_berkeley} for the PPO-05 sample.}
\label{PSDTab2}
\end{table}
\FloatBarrier

To compare the~PSD performance of the new slow liquid scintillators, a comparison between them and several LAB-based samples previously measured with the setup at the~CN accelerator was performed. Moreover, also a sample of a pseudocumene-based liquid scintillator provided by the Borexino Collaboration~(out of Borexino's central detector) was studied beforehand and also serves for comparison~(see \autoref{tab:PSD_evaluation}:~PC~+~1.5\,g/l~PPO). The results for the~PSD performance parameter~$\Delta\mu$ and the corresponding starting time of the tail integration after the time profile peak can be found in \autoref{tab:PSD_evaluation}.\\
When comparing the~LAB/PPO-based scintillators with those containing~DIN as an admixture, a clear improvement in the separation of the pulse shapes is noticeable due to the co-solvent. As expected,~PSD performance improves with increasing concentration of the fluor~(here~PPO) for both, pure LAB and LAB/DIN mixtures as solvents. The use of BPO instead of PPO has another advantage in addition to the high light yield even at low fluor concentrations. Even with~1\,g/l in a~10\%~LAB/DIN mixture (see~BPO-10 sample in \autoref{tab:PSD_evaluation}), the separation of the time profiles of the PC-based scintillator is reached. This is particularly remarkable as this scintillator is considered to feature one of the clearest pulse-shape differences in monolithic neutrino detectors to date. 

\begin{table}[h!]
    \centering
    \begin{tabular}{|c|c|c|}
         \hline 
         Sample & Maximal tail-to-total difference $\Delta \mu$ & Time after time profile peak (ns) \\ \hline 
         LAB + 0.5 g/l PPO & $0.038 \pm 0.001$ & $44.33 \pm 4.83$ \\ 
         LAB + 1.5 g/l PPO & $0.067 \pm 0.001$ & $16.33 \pm 1.29$ \\ 
         LAB + 2.0 g/l PPO & $0.106 \pm 0.004$ & $11.33 \pm 1.65$ \\ \hline 
         PPO-05 & $0.118 \pm 0.002$ & $61.83 \pm 0.96$ \\ 
         PPO-10 & $0.124 \pm 0.001$ & $45.83 \pm 0.96$ \\ 
         PPO-20 & $0.141 \pm 0.003$ & $27.17 \pm 0.59$ \\ \hline 
         BPO-10 & $0.164 \pm 0.002$ & $43.00 \pm 1.08$ \\ \hline 
         PC + 1.5 g/l PPO & $0.166\pm 0.002$ & $7.83 \pm 0.22$ \\ \hline       
    \end{tabular}
    \caption{Comparison of the tail-to-total difference between recoil proton and recoil electron events. The start time for the tail integration after the time profile peak was selected such, that it maximizes~$\Delta\mu$.}
    \label{tab:PSD_evaluation}
\end{table}

\FloatBarrier

\section{Conclusions}

Large hybrid Cherenkov/Scintillation detectors are one of the most promising approaches for the next generation of neutrino experiments. For this application blended solvent scintillation cocktails with a significantly slowed scintillation light emission compared to conventional LAB- and pseudocumene-based detection media were developed. Here,~DIN was added in concentrations of~5-20\% to conventional~LAB-based scintillation mixtures. To allow this, purification techniques based on column chromatography and fractional vacuum distillation were applied on the co-solvent greatly increasing its transparency. The resulting mixtures of highly transparent~LAB and~DIN are already suitable for use in monolithic detectors on the scale of several tons to kilotons in terms of their transparency. In addition, high light yields as required for high-resolution scintillation detectors could be achieved even with low amounts of fluor dissolved in the~LAB/DIN solvent mixture.\\ 
For selected samples detailed studies of the~Ch/S-separation by the timing of the emitted light were carried out in a table-top experiment. A purity of up to~91\% for the Cherenkov light population was achieved, while particularly long decay constants for the scintillation light emission were found~(sample~PPO-05). Furthermore, in an experiment with pulsed~QMN at the~CN accelerator of the~INFN-LNL, detailed investigations of the~PSD properties were carried out. The admixture of~DIN resulted in a significantly improved~PSD compared to pure LAB-cocktails with a similar fluor concentration.\\ 
The use of~BPO instead of the widely used~PPO has proven to be particularly promising in two ways. The resulting light yields even with small amounts of~BPO allow particularly bright but slow scintillation light emission, while the~PSD performance of the sample~BPO-10 reached the one of the pseudocmene-based scintillator used in the Borexino experiment.\\
Since the purification techniques used here have already been tested in several experiments~(e.g.~JUNO or~SNO+) on a kiloton scale and can be adapted from~LAB to~DIN from a technical point of view, the blended solvent approach has to be considered highly economical and ready for application in existing and planned large-scale neutrino detectors.\\  
Since a successful loading of LAB with the double beta-decay emitter~$^{130}$Te has already been successfully carried out \cite{Suslov}, the use of these multi-solvent scintillators is also conceivable and promising for this type of rare event search due to the improved~PID through~Ch/S separation.

\FloatBarrier

\acknowledgments

This work benefited substantially from the support and funding by the Detector Laboratory of the Cluster of Excellence~PRISMA$^+$. We are very grateful to the staff of the Detector Division around Dr. Quirin Weitzel and Dr. Anastasia Mpoukouvalas. Our special thanks go to the chemotechnical employee of the Mainz Institute for Physics Joachim Strübig, for his assistance in the fractional vacuum distillation of the co-solvent~DIN. \\
Hereby, we express our gratitude to the~JUNO collaboration for providing the~LAB samples and to the Borexino collaboration for the scintillator sample on pseudocumene basis. \\ 
Moreover, we are very grateful for the support of the Laboratori Nazionali di Legnaro~(LNL) of the Italian Istituto Nazionale di Fisica Nucleare~(INFN). For the excellent collaboration during our beamtimes we especially thank Dr. Pierfrancsco Mastinu and Dr. Elizabeth Musacchio as well as the two operators of the CN accelerator Luca Maran and Daniele Lideo.\\ 
For countless detailed and inspiring discussions we would like to thank especially Prof.~Dr.~Franz von Feilitzsch~(TUM), Dr. Brennan Hackett~(Max Planck Institute for Physics), Prof. Dr. Alberto Garfagnini (University of Padova) and Andreas Leonhardt~(TUM).\\  
The development of this novel scintillation medium was also supported by the~BMBF Collaborative Project~05H2018~-~ R\&D Detectors~(Scintillator). Moreover, this research was co-funded by the Cluster of Excellence ORIGINS which is funded by the Deutsche Forschungsgemeinschaft~(DFG, German Research Foundation) under Germany’s Excellence Strategy~–~EXC-2094~–~390783311. The work carried out in Berkeley was supported by the U.S. Department of Energy, Office of Science, Office of High Energy Physics, under Award Number DE-SC0018974.




\bibliographystyle{JHEP.bst}



\begin{thebibliography}{99}


\bibitem{Biller1}
S. Biller, Phys. Rev. D 87, 071301(R) (2013)

\bibitem{Aberle1}
C. Aberle et al, JINST 9, P06012 (2014)

\bibitem{Li1}
M. Li et al., Nuc. Instrum. and Meth. A, Volume 830 (2016)

\bibitem{Wei1}
H. Wei et al., Phys. Lett. B 769 (2017) 

\bibitem{Guo1}
Z. Guo et al., Astropart. Phys., 109 (2019)

\bibitem{Theia:2019non}
M. Askins et al., Eur. Phys. J. C, 80:416 (2020) 

\bibitem{Biller2}
S. Biller et al., Nuc. Instrum. and Meth. A, Volume 972 (2020)

\bibitem{Yeh1}
M. Yeh et al., Nuc. Instrum. and Meth. A, Volume 660 (2011)

\bibitem{Tanner1}
T. Kaptanoglu et al., Nuc. Instrum. and Meth. A, Volume 889 (2018)

\bibitem{dichroicon}
T. Kaptanoglu et al., Phys. Rev. D, vol. 101, no. 7, p. 072002 (2020)

\bibitem{Tanner2}
T. Kaptanoglu et al., JINST, vol. 14, no. 05, p. T05001 (2019)

\bibitem{ANNIE:2023yny}
Ascencio-Sosa et al., accepted by JINST, FERMILAB-PUB-23-790, arXiv: 2312.09335 (2023) 

\bibitem{Anderson:2022lbb}
T. Anderson et al., JINST 18, P02009 (2023)

\bibitem{Zhao:2023ydx}
R. Zhao et al., JINST 19, P01003 (2024)

\bibitem{Dunger:2022gif}
J. Dunger et al., Phys. Rev. D, 105, 092006 (2022)  

\bibitem{Yeh2}
M. Yeh et al., Nuc. Instrum. and Meth. A66051 (2011)

\bibitem{Lombardi1}
P. Lombardi et al., Nuc. Instrum. and Meth. A, Volume 925 (2019)

\bibitem{Franke}
S. Franke, PhD Thesis, \url{https://mediatum.ub.tum.de/doc/1453624/1453624.pdf}, (2018)

\bibitem{RAIN}
Rain Carbon Inc., Tec. Data Sheet, Ref. Nr.: 124060 (2019)

\bibitem{EJ-309}
Eljen Technology, Tec. Data Sheet Neutron/Gamma PSD Liquid Scintillator
EJ-301, EJ-309 , \url{https://eljentechnology.com/images/products/data_sheets/EJ-301_EJ-309.pdf}, (2021)

\bibitem{AberleJINST}
C. Aberle et al., JINST 7, P06008 (2012) 

\bibitem{Song}
S. Song et al., Journal of the Korean Physical Society, Volume 63 (2013)

\bibitem{BuckYeh}
Ch. Buck and M. Yeh, J. Phys. G: Nucl. Part. Phys. 43 093001 (2016)

\bibitem{Perkin}
R. Edler, LSC Appl. Note, \url{https://resources.perkinelmer.com/lab-solutions/resources/docs/app_cocktails-for-liquid-scintillation-counting-011940_01.pdf}, (2015)

\bibitem{Bonhomme}
A. Bonhomme et al., JINST 17, P11025 (2022)

\bibitem{ETEL9128B}
9128B series data sheet, \url{https://et-enterprises.com/images/data_sheets/9128B.pdf}, (2012)

\bibitem{AberleChem}
C. Aberle et al., Chemical Physics Letters 516 (2011)

\bibitem{Bollinger}
L.M. Bollinger, G.E. Thomas, Rev. Sci. Instrum. 32, 1044–1050 (1961)

\bibitem{Marrodan}
T. Marrodán Undagoitia, Rev. Sci. Instrum. 80(4):043301, (2009) 

\bibitem{HamamatsuDataSheet}
Hamamatsu H11934 Datasheet, \url{https://www.hamamatsu.com/content/dam/hamamatsu-photonics/sites/documents/99_SALES_LIBRARY/etd/R11265U_H11934_TPMH1336E.pdf}, (2019)

\bibitem{CaenV1742}
CAEN Group, Caen V1742 Switched Capacitor Digitizer, \url{https://www.caen.it/products/v1742/}, (2024)


\bibitem{CHESS}
J. Caravaca et al., Phys. Rev. C, vol. 95, no. 5, p. 055801 (2017)

\bibitem{LAPPDseparation}
T. Kaptanoglu et al., Eur. Phys. J. C, vol. 82, no. 2, p. 169 (2022)

\bibitem{SNOpScint}
M.R. Anderson et al., JINST 16, P05009 (2021)

\bibitem{Förster}
T. Förster, Discuss. Faraday Soc. 27, 7 (1959)

\bibitem{Birks}
J. B. Birks, The Theory and Practice of Scintillation Counting, Pergamon
Press, First Edition (1964)

\bibitem{Horrocks}
D. L. Horrocks, Application of Liquid Scintillation Counting, Academic Press
(1974)

\bibitem{HansPhD}
H. Steiger, PhD Thesis, \url{https://mediatum.ub.tum.de/doc/1539429/1539429.pdf}, (2020)

\bibitem{RaphaelPhD}
M. R. Stock, PhD Thesis, Techn. Univ. of Munich, (2024)

\bibitem{StockProceed}
M. R. Stock, PoS (TAUP2023) 287, (2024)

\bibitem{Lifetime}
W. Becker, The BH TCSPC Handbook. 10th edition, \url{www.becker-hickl.com}, (2023)

\bibitem{Suslov}
I.A. Suslov et al., Nuc. Instrum. and Meth. A, Volume 1040, 167131, (2022) 









\end{thebibliography}

\end{document}